\documentclass[10pt,superscriptaddress,pra,twocolumn]{revtex4}%
\usepackage{amsfonts}
\usepackage{amsmath}
\usepackage{amssymb}
\usepackage{graphicx}
\usepackage{graphics}
\usepackage[usenames]{color}%
\providecommand{\U}[1]{\protect\rule{.1in}{.1in}}

\begin{document}
\title{Green's Functions and the Adiabatic Hyperspherical Method}
\author{Seth T.~Rittenhouse}
\affiliation{Department of Physics and JILA, University of Colorado, Boulder, CO 80309}
\affiliation{ITAMP, Harvard-Smithsonian Center for Astrophysics, Cambridge, MA 02138}
\author{N.~P.~Mehta}
\affiliation{Department of Physics and JILA, University of Colorado, Boulder, CO 80309}
\affiliation{Department of Physics, Grinnell College, Grinnell, IA 50112}
\author{Chris H.~Greene}
\affiliation{Department of Physics and JILA, University of Colorado, Boulder, CO 80309}

\begin{abstract}
We address the few-body problem using the adiabatic hyperspherical
representation. A general form for the hyperangular Green's function in
$d$-dimensions is derived. The resulting Lippmann-Schwinger equation is solved
for the case of three-particles with s-wave zero-range interactions. Identical
particle symmetry is incorporated in a general and intuitive way. Complete
semi-analytic expressions for the nonadiabatic channel couplings are derived.
Finally, a model to describe the atom-loss due to three-body recombination for
a three-component fermi-gas of $^{6}$Li atoms is presented.

\end{abstract}
\maketitle

\section{Introduction}

\label{intro} In recent years, there has been extensive theoretical and
experimental interest in the area of few-body physics, most notably in the
famous effect first predicted in 1970 by Vitaly Efimov
\cite{efimov1971wbs}. Efimov studied a three boson system with
short-range two-body interactions in which each two-body system is
infinitesimally close to forming a bound state; that is, the s-wave scattering
length is infinite, or at least very large in magnitude. Quantitatively,
Efimov and later others, found that this effect is described by a simple wave
function in the hyperspherical representation
\cite{efimov1973,macek1986lbs,esry1999rta,nielsen1999ler}. A quantitative
understanding of three-body scattering \cite{esry1999rta,braaten2004edr} has
given experiments the tools to examine three-body processes in dilute gas
systems and has led to a wealth of experimental evidence for the Efimov effect
\cite{kraemer2006eeq,zaccanti2009oes,ottenstein2008cst,huckans2009tbr,gross2009ouu,pollack2009utf}%
. More recently, predictions relating to the four-body loss rate
\cite{vonstecher2009suf} have given another handle on characterizing an Efimov
resonance. The experimental realization of these predictions swiftly followed
\cite{ferlaino2009euf,pollack2009utf}.

The sticking point of the adiabatic hyperspherical method lies in solving the
adiabatic Schr\"{o}dinger equation. Often solving this $(d-1)$-dimensional
equation is as hard as solving the total $d$-dimensional Schr\"{o}dinger
equation in the first place. Having a variety of methods available is
therefore helpful. The \emph{benefit} of using the adiabatic hyperspherical
method comes from the simple final interpretation that can often be applied to
the resulting coupled set of one-dimensional equations in the
hyperradius\cite{fano1981utc}. For instance, in the three-body problem, if two
particles can form a bound state, then one of the resulting scattering
channels consists of an atom and a dimer colliding. In the adiabatic
hyperspherical method this type of fragmentation channel arises naturally as
one of the discrete solutions to the hyperangular equations. In this paper, we
derive the hyperangular Green's function for an arbitrary $d$-dimensional
system, which can then be used in a hyperangular Lippmann-Schwinger equation
to extract the adiabatic hyperradial potential curves.

This article is arranged as follows, in Section~\ref{HypGF} we derive a
general form of the hyperangular Green's function for a $d$-dimensional
system. In Section~\ref{3BZRP} the Green's function is applied to the problem
of three particles with regularized, zero-range, s-wave interactions.
Section~\ref{Li6} applies this result to the three lowest hyperfine states of
$^{6}$Li, and gives a simple description of the scaling behavior of three-body
recombination events that result in trap losses. Finally, in
Section~\ref{Summary} we summarize the results and suggest further avenues of inquiry.

\section{The hyperangular Green's function}

\label{HypGF} The adiabatic hyperspherical method has proven useful for
analyzing many few-body systems
\cite{fano1976dee,clark1980hat,Avery,zhou1993hac,lin1995hca,kokoouline2003utt}%
. The heart of this method lies in treating the overall size of the system,
the hyperradius $R$ defined by $\sqrt{\mu}R=\sqrt{\sum_{i=1}^{d}\mu_{i}%
x_{i}^{2}}$, as an adiabatic parameter. Here $\mu_{i}$ is the mass scale
associated with the $i$th cartesian coordinate and $\mu$ is the reduced mass
associated with the hyperradius. For a system of $N$ particles, $x_{i}$ denote
the $d=3N-3$ cartesian components needed to specify the relative positions of
the $N$ paticles. In this representation, the total wavefunction is written:
\begin{equation}
\Psi\left(  R,\Omega\right)  =\sum_{n}R^{-\left(  d-1\right)  /2}F_{n}\left(
R\right)  \Phi_{n}\left(  R;\Omega\right)  , \label{psitot}%
\end{equation}
where the adiabatic eigenfunctions $\Phi_{n}(R;\Omega)$ satisfy the fixed-$R$
Schr\"{o}dinger equation:
\begin{equation}
\left[  \dfrac{\hbar^{2}}{2\mu}\dfrac{\Lambda^{2}}{R^{2}}+V\left(
R,\Omega\right)  \right]  \Phi_{n}\left(  R;\Omega\right)  =u_{n}\left(
R\right)  \Phi_{n}\left(  R;\Omega\right)  , \label{Eq:Had}%
\end{equation}
Here, $\Lambda$ is the grand angular momentum operator defined by
\begin{align}
\Lambda^{2}  &  =-\sum_{i<j}\Lambda_{ij}^{2}\\
\Lambda_{ij}  &  =x_{i}\dfrac{\partial}{\partial x_{j}}-x_{j}\dfrac{\partial
}{\partial x_{i}}.
\end{align}
Inserting Eq.~(\ref{psitot}) into the full (time-independent) Schr\"{o}dinger
equation takes a $d$-dimensional partial differential equation to a set of
coupled one-dimensional differential equations:%
\begin{align}
&  \left[  -\dfrac{\hbar^{2}}{2\mu}\left(  \dfrac{d^{2}}{dR^{2}}%
-\dfrac{\left(  d-3\right)  \left(  d-1\right)  }{4R^{2}}\right)
+u_{n}\left(  R\right)  \right]  F_{n}\left(  R\right) \nonumber\\
&  -\dfrac{\hbar^{2}}{2\mu}\sum_{m}\left[  2P_{nm}\dfrac{d}{dR}+Q_{nm}\right]
F_{m}\left(  R\right)  =EF_{n}\left(  R\right)  \label{Eq:Adiab_SE}%
\end{align}
The non-adiabatic coupling matrices $P$ and $Q$ in Eq.~(\ref{Eq:Had}) are
defined as
\begin{align}
P_{mn}  &  =\left\langle \Phi_{m}\left(  R;\Omega\right)  {\Huge |}%
\dfrac{\partial}{\partial R}\Phi_{n}\left(  R;\Omega\right)  \right\rangle
,\label{Eq:Pdef}\\
Q_{mn}  &  =\left\langle \Phi_{m}\left(  R;\Omega\right)  {\Huge |}%
\dfrac{\partial^{2}}{\partial R^{2}}\Phi_{n}\left(  R;\Omega\right)
\right\rangle . \label{Eq:Qdef}%
\end{align}
The integrals in Eqs.~(\ref{Eq:Pdef}) and (\ref{Eq:Qdef}) are taken only over
the $d-1$ hyperangles collectively denoted $\Omega$.

Approximate solutions can be found by solving the uncoupled system of
equations, which are referred to as the adiabatic approximation:%
\begin{align}
&  \left[  -\dfrac{\hbar^{2}}{2\mu}\dfrac{d^{2}}{dR^{2}}+\dfrac{\hbar^{2}%
}{2\mu}\dfrac{\left(  d-3\right)  \left(  d-1\right)  }{4R^{2}}\right.
\nonumber\\
&  -\left.  \dfrac{\hbar^{2}}{2\mu}Q_{nn}+u_{n}\left(  R\right)  \right]
F_{n}\left(  R\right)  =EF_{n}\left(  R\right)  \label{uncouprad}%
\end{align}
The ground state eigenenergy Eq.~(\ref{uncouprad}) is a variational upper
bound to the exact ground state energy from Eq.~(\ref{Eq:Adiab_SE}). Another
variant of this method is frequently denoted the Born-Oppenheimer
approximation, with the diagonal correction $-\dfrac{\hbar^{2}}{2\mu}%
Q_{nn}(R)$ to the potential $u_{n}(R)$ omitted. These two approximations will
be the main focus of this paper, while the non-adiabatic couplings $P_{nm}(R)$
will be used to describe Landau-Zener-Stueckelberg transitions between the
different hyperradial channels $u_{n}\left(  R\right)  $. Once the adiabatic
potentials have been found, much of the intuition of simple one dimensional
Schr\"{o}dinger quantum mechanics can be brought to bear upon the problem.
Unfortunately, obtaining these potentials can be prohibitively difficult in
many problems; the development of efficient ways to calculate them is desirable.

This section derives the free space hyperangular Green's function for an
arbitrary $d$ dimensional space such as an $N$-body system with $d=3\left(
N-1\right)  $ with the center of mass coordinate removed. This Green's
function can then be used to recast equation Eq.~(\ref{Eq:Had}) into an
integral Lippmann-Schwinger (LS) equation. The $d$ dimensional Laplacian
written in hyperspherical coordinates is given in Ref. \cite{Avery} as
\begin{equation}
\nabla^{2}=\dfrac{1}{R^{\left(  d-1\right)  /2}}\dfrac{\partial^{2}}{\partial
R^{2}}R^{\left(  d-1\right)  /2}-\dfrac{\left(  d-1\right)  \left(
d-3\right)  }{4R^{2}}-\dfrac{\mathbf{\Lambda}^{2}}{R^{2}} \label{Laplacian}%
\end{equation}
The hyperangular Green's function is given as the solution to
\begin{equation}
\left[  \mathbf{\Lambda}^{2}-\nu\left(  \nu+d-2\right)  \right]  G^{\nu
}\left(  \Omega,\Omega^{\prime}\right)  =\delta^{d}\left(  \Omega
-\Omega^{\prime}\right)  . \label{GFEq}%
\end{equation}
Here $\Omega$ stands for the $d-1$ hyperangular coordinates needed to describe
the surface of a $d$ dimensional hypersphere and $\delta^{d}\left(
\Omega-\Omega^{\prime}\right)  $ is the Dirac $\delta$-function in the
hyperangular coordinates, i.e. $\delta\left(  \Omega-\Omega^{\prime}\right)
=0$ if $\Omega\neq\Omega^{\prime}$ and $\int\delta^{d}\left(  \Omega
-\Omega^{\prime}\right)  d\Omega=1$. The Green's function can be found in
several forms, including the full hyperspherical harmonic expansion
\cite{fabredelaripelle1993gfa}, and has been given in closed form given by
Szmytkowski in Ref. \cite{szmytkowski2006cfg}.

The simplest derivation of the Green's function relies on the completeness of
hyperspherical harmonics:%
\begin{equation}
\sum_{\lambda\mu}Y_{\lambda\mu}^{\ast}\left(  \Omega^{\prime}\right)
Y_{\lambda\mu}\left(  \Omega\right)  =\delta^{d}\left(  \Omega-\Omega^{\prime
}\right)  . \label{Eq:HH_completeness}%
\end{equation}
The function $Y_{\lambda\mu}$ is the solution to the eigenvalue equation%
\begin{equation}
\mathbf{\Lambda}^{2}Y_{\lambda\mu}\left(  \Omega\right)  =\lambda\left(
\lambda+d-2\right)  Y_{\lambda\mu}\left(  \Omega\right)  .
\label{Eq:hyperharm_eigeq}%
\end{equation}
Here $\lambda$ is the hyperangular momentum quantum number, and $\mu$
enumerates the degenerate states. These functions are generally expressed as
products of Jacobi polynomials for any number of dimensions, and are
thoroughly described by a number of authors (See Refs.
\cite{Avery,SmirnovShitikova} for some examples). They are simply an extension
of normal spherical harmonics to higher dimension.

Equation (\ref{Eq:HH_completeness}) can be used in conjunction with
Eq.~(\ref{Eq:hyperharm_eigeq}) to find $G^{\nu}(\Omega,\Omega^{\prime})$
\cite{fabredelaripelle1993gfa}:%
\begin{equation}
G^{\nu}\left(  \Omega,\Omega^{\prime}\right)  =\sum_{\lambda\mu}%
\dfrac{Y_{\lambda\mu}^{\ast}\left(  \Omega^{\prime}\right)  Y_{\lambda\mu
}\left(  \Omega\right)  }{\lambda\left(  \lambda+d-2\right)  -\nu\left(
\nu+d-2\right)  }. \label{Eq:GF_spectral}%
\end{equation}
Unfortunately, eigenfunction expansions of Green's function often have slow
convergence with respect to the number of eigenfunctions, making them
unsuitable for numerical calculations. The closed form of the Green's function
from Ref. \cite{szmytkowski2006cfg} is given as%
\begin{equation}
G^{\nu}\left(  \Omega,\Omega^{\prime}\right)  =\dfrac{-\pi}{\left(
d-2\right)  S_{d}\sin\pi\nu}C_{\nu}^{\left(  d-2\right)  /2}\left(  -\hat
{R}\cdot\hat{R}^{\prime}\right)  , \label{Eq:GF_closedform}%
\end{equation}
where $C_{\nu}^{\alpha}$ is a Gegenbauer function, $S_{d}$ is the surface area
of the $d$-dimensional unit hypersphere: $S_{d}=\int d\Omega=2\pi^{d/2}%
/\Gamma\left(  d/2\right)  $, and $\hat{R}\cdot\hat{R}^{\prime}$ is the cosine
of the angle between the two normalized hypervectors $\hat{R}$ and $\hat
{R}^{\prime}$. Here $\nu$ is defined by Eq.~(\ref{GFEq}). While
Eq.~\ref{Eq:GF_closedform} has a pleasing, compact form it is often divergent
at critical points. For instance if $\nu$ is non-integer valued, then $G^{\nu
}\left(  \Omega,\Omega^{\prime}\right)  $ diverges as $\hat{R}\cdot\hat
{R}^{\prime}\rightarrow1$.

For these reasons, it is convenient to find a third form of the Green's
function. The first step in this derivation relies on the division of the
total $d$ dimensional space into two subspaces. For the purposes of this work,
we will assume that the dimension of the two subspaces are both greater than
$2$: i.e. $d_{1},d_{2}\geq2$. The two subspaces are each described by
sub-hyperspherical coordinates. The the two resulting sub-hyperradii can then
be related to the total hyperradius as%
\begin{align}
R_{1}  &  =R\sin\alpha,\label{subradiuscoor}\\
R_{2}  &  =R\cos\alpha,\nonumber\\
0  &  \leq\alpha\leq\pi/2.\nonumber
\end{align}
Reference \cite{SmirnovShitikova} details how the hyperangular momentum can be
written in terms of the sub-hyperangular momenta as in
Eq.~(\ref{Eq:Hyperangmoment2}). With the following definitions:
\begin{align}
\mathbf{\Lambda}_{1}^{2}Y_{\lambda_{1}\mu_{1}}\left(  \Omega_{1}\right)   &
=\lambda_{1}\left(  \lambda_{1}+d_{1}-2\right)  Y_{\lambda_{1}\mu_{1}}\left(
\Omega_{1}\right)  ,\label{Eq:Subhyperharm}\\
\mathbf{\Lambda}_{2}^{2}Y_{\lambda_{2}\mu_{2}}\left(  \Omega_{2}\right)   &
=\lambda_{2}\left(  \lambda_{2}+d_{2}-2\right)  Y_{\lambda_{2}\mu_{2}}\left(
\Omega_{2}\right)  ,\nonumber
\end{align}
the Green's function can be expanded using the completeness of the
sub-hyperspherical harmonics [viz. Eq.~(\ref{Eq:Greenseq})]. Substituting the
expansion Eq.~(\ref{Eq:Greenseq}) into Eq.~(\ref{GFEq}), we find that the
latter is satisfied if and only if Eq.~(\ref{littlegreen}) is satisfied;
$\delta\left(  \alpha-\alpha^{\prime}\right)  $ is a Dirac $\delta$-function
and the denominator on the left hand side of Eq.~(\ref{littlegreen}) arises
from the hyperangular volume element associated with the angle $\alpha$ (See
Refs. \cite{Avery,szmytkowski2006cfg} for details).
\begin{widetext}
\begin{align}
\mathbf{\Lambda}^{2} &= \dfrac{-1}{\left(  \sin\alpha\right)^{\left(d_{1}-1\right)  /2}\left(  \cos\alpha\right)^{\left(  d_{2}-1\right)  /2}}
\dfrac{\partial^{2}}{\partial\alpha^{2}}\left(  \sin\alpha\right)  ^{\left(d_{1}-1\right)  /2}\left(  \cos\alpha\right)  ^{\left(  d_{2}-1\right)/2}\nonumber\\
& +\dfrac{\mathbf{\Lambda}_{1}^{2}+\left(  d_{1}-1\right)  \left(d_{1}-3\right)  /4}{\sin^{2}\alpha}+\dfrac{\mathbf{\Lambda}_{2}^{2} +\left(d_{2}-1\right)  \left(  d_{2}-3\right)  /4}{\cos^{2}\alpha} - \dfrac{\left(d-1\right)  \left(  d-3\right)  +1}{4}.\label{Eq:Hyperangmoment2}
\end{align}
\begin{equation}
G\left(  \Omega,\Omega^{\prime}\right)  =\sum_{\lambda_{1}\mu_{1}}%
\sum_{\lambda_{2}\mu_{2}}g\left(  \alpha,\alpha^{\prime}\right)
Y_{\lambda_{1}\mu_{1}}^{\ast}\left(  \Omega_{1}^{\prime}\right)
Y_{\lambda_{1}\mu_{1}}\left(  \Omega_{1}\right)  Y_{\lambda_{2}\mu_{2}}^{\ast
}\left(  \Omega_{2}^{\prime}\right)  Y_{\lambda_{2}\mu_{2}}\left(  \Omega
_{2}\right)  , \label{Eq:Greenseq}%
\end{equation}
\begin{align}
\dfrac{\delta\left(  \alpha-\alpha^{\prime}\right)  }{\left(  \sin
\alpha\right)  ^{d_{1}-1}\left(  \cos\alpha\right)  ^{d_{2}-1}}=  &  \left[
\dfrac{-1}{\left(  \sin\alpha\right)  ^{d_{1}-1}\left(  \cos\alpha\right)
^{d_{2}-1}}\dfrac{\partial}{\partial\alpha}\left(  \sin\alpha\right)
^{d_{1}-1}\left(  \cos\alpha\right)  ^{d_{2}-1}\dfrac{\partial}{\partial
\alpha}\right. \nonumber \\
&  +\dfrac{\mathbf{\lambda}_{1}\left(  \lambda_{1}+d_{1}-2\right)  }{\sin
^{2}\alpha}+\left.  \dfrac{\mathbf{\lambda}_{2}\left(  \lambda_{2}%
+d_{2}-2\right)  }{\cos^{2}\alpha}-\nu\left(  \nu+d-2\right)  \right]
g_{\lambda_{1},\lambda_{2}}^{d_{1},d_{2}}\left(  \nu;\alpha,\alpha^{\prime
}\right)  .\label{littlegreen}
\end{align}
The general one-dimensional Green's function for any differential equation of
Sturm-Liouville form Eq.~(\ref{littlegreen}) is:
\begin{equation}
g_{\lambda_{1},\lambda_{2}}^{d_{1},d_{2}}\left(  \nu;\alpha,\alpha^{\prime
}\right)  =\dfrac{-f_{\lambda_{1}\lambda_{2}\nu}^{+}\left(  \alpha_{<}\right)
f_{\lambda_{1}\lambda_{2}\nu}^{-}\left(  \alpha_{>}\right)  }{\left(
\sin\alpha\right)  ^{d_{1}-1}\left(  \cos\alpha\right)  ^{d_{2}-1}W\left[
f^{+},f^{-}\right]  }, \label{GFWronsk}%
\end{equation}
where $W\left[  f^{+},f^{-}\right]  =f^{+}f^{-\prime}-f^{-}f^{+\prime}$ is the
Wronskian~\cite{jackson1999classical} and $\alpha_{<\left(  >\right)  }%
=\min\left(  \alpha,\alpha^{\prime}\right)  $ ($\max\left(  \alpha
,\alpha^{\prime}\right)  $). The functions $f^{+}\left(  \alpha\right)  $ and
$f^{-}\left(  \alpha\right)  $ are regular at $\alpha=0$ and $\alpha=\pi/2$
respectively and satisfy the homogeneous version of Eq.~(\ref{littlegreen}).
The solutions $f^{+}\left(  \alpha\right)  $ and $f^{-}\left(  \alpha\right)
$ are given in Ref.~\cite{abramowitz1965hmf} as
\begin{align}
f_{\lambda_{1}\lambda_{2}\nu}^{\left(  \pm\right)  }\left(  \alpha\right)   &
=\left(  \sin^{\lambda_{1}}\alpha\cos^{\lambda_{2}}\alpha{}\right)  {}%
_{2}F_{1}\left(  \tfrac{\lambda_{1}+\lambda_{2}-\nu}{2},\tfrac{\nu+\lambda
_{1}+\lambda_{2}+d-2}{2};\tfrac{2\lambda_{\pm}+d_{\pm}}{2};\tfrac{1\mp
\cos2\alpha}{2}\right)  ,\label{Eq:PMEq}\\
W\left[  f_{\lambda_{1}\lambda_{2}\nu}^{+},f_{\lambda_{1}\lambda_{2}\nu}%
^{-}\right]   &  =\dfrac{-2\Gamma\left(  \tfrac{2\lambda_{1}+d_{1}}{2}\right)
\Gamma\left(  \tfrac{2\lambda_{2}+d_{2}}{2}\right)  }{\left(  \sin
\alpha\right)  ^{d_{1}-1}\left(  \cos\alpha\right)  ^{d_{2}-1}\Gamma\left(
\tfrac{\nu+\lambda_{1}+\lambda_{2}+d-2}{2}\right)  \Gamma\left(
\tfrac{\lambda_{1}+\lambda_{2}-\nu}{2}\right)  }, \label{Eq:Wronsk}%
\end{align}
\end{widetext}
where $_{2}F_{1}\left(  a,b;c,x\right)  $ is a hypergeometric function,
$\lambda_{+}=\lambda_{1}$, $d_{+}=d_{1}$, $\lambda_{-}=\lambda_{2}$, and
$d_{-}=d_{2}$.

\section{The Three-Body Problem with Zero-Range Interactions}

\label{3BZRP} In this section we show the utility of the Green's function
developed in the previous section by applying it to the three body problem
with regularized, zero-range, s-wave, pseudo-potential interactions. This
problem has been well studied by a variety of sources
\cite{nielsen2001tbp,esry1999rta,dincao2005sls,Braaten2006physrep}. The full
Hamiltonian for the untrapped system is given by%
\begin{equation}
H_{tot}=\sum_{i=1}^{3}-\dfrac{\hbar^{2}}{2m_{i}}\nabla_{i}^{2}+\sum
_{i>j}V_{ij}\left(  r_{ij}\right)  , \label{Eq:3BHam_1}%
\end{equation}
where $\vec{r}_{i}$ is the position of the $i$th particle, and $\nabla_{i}%
^{2}$ is the Laplacian for $\vec{r}_{i}$ . The interaction is given by%
\begin{equation}
V_{ij}\left(  r_{ij}\right)  =\dfrac{4\pi\hbar^{2}a_{ij}}{2\mu_{ij}}%
\delta^{\left(  3\right)  }\left(  \vec{r}_{ij}\right)  \dfrac{\partial
}{\partial r_{ij}}r_{ij}, \label{Pseudopot}%
\end{equation}
where $a_{ij}$ is the s-wave scattering length between particles $i$ and $j$
and $\mu_{ij}$ is the two body reduced mass, $\mu_{ij}=m_{i}m_{j}/\left(
m_{i}+m_{j}\right)  $. The pseudo-potential defined in this way applies the
Bethe-Peierls boundary condition to the two-body wave function as
$r\rightarrow0$, i.e. $\psi\left(  r\right)  \rightarrow\left(  1-a_{ij}%
/r\right)  C$ for some constant $C$~\cite{ross1961mer}. The center of mass can
be removed from this system by converting to a system of Jacobi vectors.
Jacobi vectors are created for this system by considering the separation
vector between two of the three particles and then a second vector from the
center of mass of that two body system to the third. The final vector is then
just the center of mass coordinate. The choice of Jacobi vectors is not
unique. Here we will need to consider three different Jacobi coordinate
parametrizations each of which is convenient for describing one of the three
possible two-body interactions $V\left(  r_{ij}\right)  .$ In the
\textquotedblleft odd-man-out" notation these are given by%
\begin{align}
\vec{\rho}_{1}^{\left(  k\right)  }  &  =\left(  \vec{r}_{i}-\vec{r}%
_{j}\right)  /d_{k},\nonumber\\
\vec{\rho}_{2}^{\left(  k\right)  }  &  =d_{k}\left(  \dfrac{m_{i}\vec{r}%
_{i}+m_{j}\vec{r}_{j}}{m_{i}+m_{j}}-\vec{r}_{k}\right)  ,\label{Jaccorrds}\\
\vec{r}_{CM}  &  =\dfrac{\left(  m_{1}\vec{r}_{1}+m_{2}\vec{r}_{2}+m_{3}%
\vec{r}_{3}\right)  }{m_{1}+m_{2}+m_{3}},\nonumber\\
d_{k}^{2}  &  =\dfrac{\left(  m_{k}/\mu\right)  \left(  m_{i}+m_{j}\right)
}{m_{1}+m_{2}+m_{3}},\nonumber
\end{align}
where $\mu$ is the three-body reduced mass:%
\begin{equation}
\mu=\sqrt{\dfrac{m_{1}m_{2}m_{3}}{m_{1}+m_{2}+m_{3}}}.
\end{equation}

The total Hamiltonian can be rewritten in terms of the Jacobi coordinates and
the center of mass as%
\begin{align}
H_{tot}  &  =H+H_{CM},\label{Eq:Hamtot2}\\
H_{CM}  &  =\dfrac{-\hbar^{2}}{2M}\nabla_{CM}^{2},\nonumber\\
H  &  =-\dfrac{\hbar^{2}}{2\mu}\sum_{i=1}^{2}\nabla_{\rho_{i}}^{2}+\sum
_{i>j}V_{ij}\left(  r_{ij}\right)  .\nonumber
\end{align}
Transforming the Jacobi coordinate piece of the Hamiltonian in
Eq.~(\ref{Eq:Hamtot2}) into hyperspherical coordinates using
Eqs.~(\ref{Laplacian}) and (\ref{Jaccorrds}) yields%
\begin{equation}
H=-\dfrac{\hbar^{2}}{2\mu}\dfrac{1}{R^{5/2}}\dfrac{\partial^{2}}{\partial
R^{2}}R^{5/2}+\dfrac{15\hbar^{2}}{8\mu R^{2}}+\dfrac{\hbar^{2}\Lambda^{2}%
}{2\mu R^{2}}+\sum_{i<j}V_{ij}\left(  d_{k}\rho_{1}^{\left(  k\right)
}\right)  . \label{FullSE}%
\end{equation}
To apply the adiabatic hyperspherical formulation, the hyperangular adiabatic
Schr\"{o}dinger equation must be solved:%
\begin{equation}
\left[  \Lambda^{2}+\dfrac{2\mu R^{2}}{\hbar^{2}}\sum_{i<j}V_{ij}\left(
d_{k}\rho_{1}^{\left(  k\right)  }\right)  -\nu\left(  \nu+4\right)  \right]
\Phi\left(  R;\Omega\right)  =0.
\end{equation}
This can now be accomplished with the use of the hyperangular Green's
function, Eq.~(\ref{Eq:Greenseq}), in the Lippmann-Schwinger (LS) equation,%
\begin{align}
\Phi\left(  R;\Omega\right)   &  =-\dfrac{2\mu R^{2}}{\hbar^{2}}\int
d\Omega^{\prime}G^{\nu}\left(  \Omega,\Omega^{\prime}\right) \label{LSEq}\\
&  \times\left[  \sum_{i<j}V_{ij}\left(  d_{k}\rho_{1}^{\left(  k\right)
\prime}\right)  \right]  \Phi\left(  R;\Omega^{\prime}\right)  ,\nonumber
\end{align}
where $\vec{\rho}_{1}^{\left(  k\right)  \prime}$ is the $k$th Jacobi vector
parametrized by $\left\{  R,\Omega^{\prime}\right\}  $. Because the system has
been constrained to have a constant hyperradius, this is effectively a bound
state problem; note that Eq.~(\ref{LSEq}) has been assumed here to have no
noninteracting solution at the chosen value of $\nu$. The hyperradial
Hamiltonian from Eq.~(\ref{Eq:Adiab_SE}) in the absence of the non-adiabatic
couplings $P$ and $Q$ is given in terms of the hyperangular eigenvalue $\nu$
as%
\begin{align}
H_{R}  &  =\dfrac{-\hbar^{2}}{2\mu}\dfrac{\partial^{2}}{\partial R^{2}}%
+U_{n}\left(  R\right)  ,\label{EffH}\\
U_{n}\left(  R\right)   &  =\dfrac{\hbar^{2}}{2\mu}\left[  \dfrac{\left(
\nu_{n}+2\right)  ^{2}-1/4}{R^{2}}-Q_{nn}\left(  R\right)  \right]  .\nonumber
\end{align}

To evaluate the integrals in the LS equation, the Green's function from
Eq.~(\ref{Eq:Greenseq}) is expressed in terms of the appropriate Jacobi
coordinate set for each interaction term in the sum, with the hyperangles
defined as%
\begin{equation}
\Omega^{\left(  k\right)  }=\left\{  \omega_{1}^{\left(  k\right)  }%
,\omega_{2}^{\left(  k\right)  },\alpha^{\left(  k\right)  }\right\}  ,
\label{hyperangs3b}%
\end{equation}
where $\omega_{i}^{\left(  k\right)  }$ represent the spherical polar angular
coordinates for $\vec{\rho}_{i}^{\left(  k\right)  }$. The remaining
hyperangle $\alpha^{\left(  k\right)  }$ is defined as in
Eq.~(\ref{subradiuscoor}), i.e.%
\begin{align}
\rho_{1}^{\left(  k\right)  }  &  =R\sin\alpha^{\left(  k\right)
},\label{alphdef2}\\
\rho_{2}^{\left(  k\right)  }  &  =R\cos\alpha^{\left(  k\right)  }.\nonumber
\end{align}
With this choice of hyperangles, it is clear that $d_{1}=d_{2}=3$ and the
hyperspherical sub-harmonics $Y_{\lambda_{i}\mu_{i}}^{\left(  i\right)
}\left(  \Omega_{i}\right)  $ in Eq.~(\ref{Eq:Greenseq}) reduce to normal
spherical harmonics $y_{L_{i}M_{i}}\left(  \omega_{i}\right)  $.

The $\delta$-function implies that the Bethe-Peierls two-body boundary
condition for each two-body interaction can be applied and the third particle
can be considered to be far away, i.e.%
\begin{equation}
\lim\limits_{\rho_{1}^{\left(  k\right)  }\rightarrow0}\Phi\left(
R;\Omega\right)  =\left(  1-\dfrac{a^{\left(  k\right)  }}{d_{k}\rho
_{1}^{\left(  k\right)  }}\right)  y_{LM}\left(  \omega_{2}^{\left(  k\right)
}\right)  C_{LM}^{\left(  k\right)  }. \label{3BBC}%
\end{equation}
Here $y_{LM}$ is a spherical harmonic describing the free space behavior in
$\omega_{2}^{\left(  k\right)  }$ and it carries the total angular momentum of
the system. The superscript $k$ again indicates the odd man out notation. This
gives the values of the sub-hyperangular momentum quantum numbers in the $k$
Jacobi coordinate system as $\lambda_{1}=0$ and $\lambda_{2}=L$, which
accounts for the s-wave interaction and the total angular momentum $L$.
Inserting Eq.~(\ref{3BBC}) into Eq.~(\ref{LSEq}) gives the hyperangular
eigenfunction,%
\begin{widetext}
\begin{align}
\Phi\left(  R;\Omega\right)  =  &  \dfrac{2\mu}{R}\sum_{k}\dfrac{a^{\left(
k\right)  }}{2\mu_{k}d_{k}^{3}}N_{L\nu}C_{LM}^{\left(  k\right)  }%
y_{LM}\left(  \omega_{2}^{\left(  k\right)  }\right)  f_{0L\nu}^{-}\left(
\alpha^{\left(  k\right)  }\right)  ,\label{Hyperangwf}\\
N_{L\nu}=  &  \dfrac{-\Gamma\left(  \dfrac{L-\nu}{2}\right)  \Gamma\left(
\dfrac{L+\nu+4}{2}\right)  }{\sqrt{\pi}\Gamma\left(  L+\dfrac{3}{2}\right)
},\nonumber
\end{align}
where $\mu_{k}$ is the two-body reduced mass labeled in the odd man out
notation and the orthonormality of spherical harmonics has been used to
evaluate the $\omega_{1}^{\left(  k\right)  \prime}$ and $\omega_{2}^{\left(
k\right)  \prime}$ integrals. The $\delta$-function in Eq.~(\ref{Pseudopot})
implies that the integral in $\alpha^{\left(  k\right)  \prime}$ can be
accomplished by evaluating at $\alpha_{<}^{\left(  k\right)  }=\alpha^{\left(
k\right)  \prime}=0.$

The analytic equation for the hyperangular eigenfunction in
Eq.~(\ref{Hyperangwf}) is not very useful without knowing the hyperangular
eigenvalue $\nu(R)$. To obtain an equation for $\nu$ the boundary condition
given in Eq.~(\ref{3BBC}) must be applied again, i.e.
\begin{align}
y_{LM}\left(  \omega_{2}^{\left(  k^{\prime}\right)  }\right)  C_{LM}^{\left(
k^{\prime}\right)  }  &  =\lim\limits_{\alpha^{\left(  k^{\prime}\right)
}\rightarrow0}\dfrac{\partial}{\partial\alpha^{\left(  k^{\prime}\right)  }%
}\alpha^{\left(  k^{\prime}\right)  }\Phi\left(  R;\Omega^{\left(  k^{\prime
}\right)  }\right) \nonumber\\
&  =\dfrac{2\mu}{R}\sum_{k}\dfrac{a^{\left(  k\right)  }}{2\mu_{k}d_{k}^{3}%
}N_{L\nu}C_{LM}^{\left(  k\right)  }\lim\limits_{\alpha^{\left(  k^{\prime
}\right)  }\rightarrow0}\dfrac{\partial}{\partial\alpha^{\left(  k^{\prime
}\right)  }}\alpha^{\left(  k^{\prime}\right)  }\left[  f_{0L\nu}^{-}\left(
\alpha^{\left(  k\right)  }\right)  y_{LM}\left(  \omega_{2}^{\left(
k\right)  }\right)  \right]  , \label{mateq}%
\end{align}
\end{widetext}
To evaluate the limit on the right hand side of this, we must determine the
values of the $k\neq k^{\prime}$ Jacobi coordinates in the limit $\rho
_{1}^{\left(  k^{\prime}\right)  }\rightarrow0$. Equations~(\ref{Jaccorrds})
and (\ref{alphdef2}) give, for $k\neq k^{\prime}$,%
\begin{align}
&  \lim\limits_{\alpha^{\left(  k^{\prime}\right)  }\rightarrow0}%
\alpha^{\left(  k\right)  }=\beta_{kk^{\prime}}=\arctan\left[  \dfrac{\left(
m_{1}+m_{2}+m_{3}\right)  \mu}{m_{k}m_{k^{\prime}}}\right] \\
&  \lim\limits_{\alpha^{\left(  k^{\prime}\right)  }\rightarrow0}\vec{\rho
}_{2}^{\left(  k\right)  }\propto-\vec{\rho}_{2}^{\left(  k^{\prime}\right)
}. \label{Eq:Jacveclim}%
\end{align}
Note that if $f^{-}$ is regular at $\beta_{kk^{\prime}}$, then
\begin{equation}
\lim\limits_{\alpha^{\left(  k^{\prime}\right)  }\rightarrow0}\dfrac{\partial
}{\partial\alpha^{\left(  k^{\prime}\right)  }}\alpha^{\left(  k^{\prime
}\right)  }f_{0L\nu}^{-}\left(  \alpha^{\left(  k\right)  }\right)
\rightarrow f_{0L\nu}^{-}\left(  \beta_{kk^{\prime}}\right)  .
\end{equation}
Using this and evaluating the limits in Eq.~(\ref{mateq}) yields a matrix
equation for $C_{LM}^{\left(  k\right)  }$:%
\begin{align}
C_{LM}^{\left(  k^{\prime}\right)  }  &  =\sum_{k}M_{k^{\prime}k}^{L\nu}%
C_{LM}^{\left(  k\right)  },\label{TrancendME}\\
M_{k^{\prime}k}^{L\nu}  &  =\left\{
\begin{array}
[c]{ccc}%
\dfrac{2\mu}{R}\dfrac{2\Gamma\left(  \tfrac{L-\nu}{2}\right)  \Gamma\left(
\tfrac{\nu+L+4}{2}\right)  }{\Gamma\left(  \tfrac{L-\nu-1}{2}\right)
\Gamma\left(  \tfrac{L+\nu+3}{2}\right)  }\dfrac{a^{\left(  k^{\prime}\right)
}}{2\mu_{k}d_{k}^{3}} & \text{for} & k=k^{\prime}\\
\left(  -1\right)  ^{L}\dfrac{2\mu}{R}N_{L\nu}\dfrac{a^{\left(  k\right)  }%
}{2\mu_{k}d_{k}^{3}}f_{0L\nu}^{-}\left(  \beta_{kk^{\prime}}\right)  &
\text{for} & k\neq k^{\prime}%
\end{array}
\right.  .\nonumber
\end{align}
The hyperangular eigenvalue, $\nu$, is found by solving the closed form
transcendental equation,
\begin{equation}
\det\left(  \mathbf{M}-\mathbf{1}\right)  =0, \label{traneq}%
\end{equation}
for any given total angular momentum $L$, any set of s-wave scattering lengths
$a^{\left(  k\right)  }$ and arbitrary masses.

\subsection{Imposing symmetry}

\label{Sym} The hyperangular eigenvalues for the general three-body problem
with arbitrary exchange symmetry can be found by solving the transcendental
equation~(\ref{traneq}), but the system can be simplified by considering
different permutation symmetries and imposing those symmetries on the boundary
conditions $C_{LM}^{\left(  k\right)  }$. For example, if the particles in
question are identical bosons, permutation cannot have any effect on the wave
function. Thus, if two particles are exchanged in the two-body subsystem, the
boundary condition must remain the same, i.e. $C_{LM}^{\left(  1\right)
}=C_{LM}^{\left(  2\right)  }=C_{LM}^{\left(  3\right)  }=C_{LM}$ and
$a^{\left(  1\right)  }=a^{\left(  2\right)  }=a^{\left(  3\right)  }=a$. A
complete list of the possible exchange symmetries is given in
Table~\ref{Symmtab}.

To illustrate this post-symmetrization, we apply the identical boson symmetry
with $L=0$ to Eq.~(\ref{TrancendME}) resulting in the well known
transcendental equation for $\nu$
\cite{efimov1970ela,nielsen2001tbp,esry1999rta,dincao2005sls,Braaten2006physrep}%
,%
\begin{equation}
\dfrac{R}{a}=\dfrac{-3^{1/4}\left[  \dfrac{8}{\sqrt{3}}\sin\left(  \dfrac
{\pi\left(  \nu+2\right)  }{6}\right)  -\left(  \nu+2\right)  \cos\left(
\dfrac{\pi\left(  \nu+2\right)  }{2}\right)  \right]  }{\sqrt{2}\sin\left(
\dfrac{\pi\left(  \nu+2\right)  }{2}\right)  }. \label{BBBeq}%
\end{equation}
In the limit where $R\ll\left\vert a\right\vert $ the first solution to this
transcendental equation gives a super-critical attractive $1/R^{2}$ effective
potential,%
\begin{align}
U\left(  R\right)   &  =\dfrac{\hbar^{2}}{2\mu}\dfrac{-s_{0}^{2}-1/4}{R^{2}%
},\label{Eq:BBB_pot}\\
s_{0}  &  =1.00624.\nonumber
\end{align}
This attractive potential is the source of the famous Efimov effect, where an
effective attractive dipole-type potential supports an infinite set of
three-body bound states that accumulate at the non-interacting three-body
threshold, $E=0$. \begin{table}[h]
\begin{center}%
\begin{tabular}
[c]{|l|lll|}\hline
X$_{1}$X$_{2}$X$_{3}$ & $C^{\left(  1\right)  }$ & $C^{\left(  2\right)  }$ &
$C^{\left(  3\right)  }$\\\hline
BBB & $C$ & $C$ & $C$\\
BBX & $C_{1}$ & $C_{1}$ & $C_{2}$\\
FFX & $C$ & $-C$ & $0$\\\hline
\end{tabular}
\end{center}
\caption{The possible permutation symmetries that may be imposed on the three
body system with s-wave interactions are given with the appropriate boundary
conditions. B stands for a boson, F for a fermion and X for a distinguishable
particle with an arbitrary mass.}%
\label{Symmtab}%
\end{table}

\subsection{Non-adiabatic couplings}

\label{Couplings} As in any adiabatic treatment, the effective hyperradial
potentials are coupled by non-adiabatic terms that arise from the hyperradial
dependence of the hyperangular channel functions. These couplings come in the
form of the $\mathbf{P}$- and $\mathbf{Q}$-matrices in Eq.~(\ref{Eq:Adiab_SE}%
). To find the non-adiabatic coupling matrices, we apply the methods of Ref.
\cite{kartavtsev2007let}. The details of the derivation are shown in
Appendix~A, the result of which gives the semi-analytic expressions for the
matrix elements $P_{mn}$:%
\begin{align}
P_{mn}  &  =\dfrac{\sum_{k}C_{m}^{\left(  k\right)  }C_{n}^{\left(  k\right)
}\dfrac{a^{\left(  k\right)  }}{d_{k}R^{2}}}{\left(  \varepsilon
_{m}-\varepsilon_{n}\right)  }\text{ for }n\neq m\label{Eq:Pmat_elem}\\
-\varepsilon_{n}^{\prime}  &  =\sum_{k}\left(  C_{n}^{\left(  k\right)
}\right)  ^{2}\dfrac{a^{\left(  k\right)  }}{d_{k}R^{2}}. \label{Eq:Norm_cond}%
\end{align}
Here, for notational simplicity, we have set $\varepsilon_{n}=\left(  \nu
_{n}+2\right)  ^{2}$ and all primes indicate a derivative with respect to $R$
(e.g. $\varepsilon_{n}^{\prime}=d\varepsilon_{n}/dR$). Because the
hyperangular eigenfunctions are orthonormal, the diagonal part of the $P$
matrix is zero, i.e. $P_{nn}=\dfrac{1}{2}\dfrac{\partial}{\partial
R}\left\langle \Phi_{n}|\Phi_{n}\right\rangle =0$. Equation
(\ref{Eq:Norm_cond}) gives the normalization condition for $\Phi_{n}$, with an
overall phase that is free. This overall phase is chosen here so that
$\sum_{k}C_{n}^{\left(  k\right)  }$ is positive. A similar derivation
provides the matrix elements $Q_{mn}$:%
\begin{align}
Q_{mn}  &  =\delta_{mn}\left(  \dfrac{\varepsilon_{n}^{\prime}+R\varepsilon
_{n}^{\prime\prime}+A_{n}}{R^{2}\varepsilon_{n}^{\prime}}+\dfrac
{\varepsilon_{n}^{\prime\prime\prime}}{6\varepsilon_{n}^{\prime}}\right)
\nonumber\\
&  +2\left(  1-\delta_{mn}\right)  \dfrac{\varepsilon_{n}^{\prime}%
P_{mn}+B_{mn}}{\left(  \varepsilon_{m}-\varepsilon_{n}\right)  }%
,\label{Qmat1}\\
A_{n}  &  =\sum_{k}\dfrac{a^{\left(  k\right)  }}{d_{k}}\left[  \left(
C_{n}^{\left(  k\right)  }\right)  ^{\prime}\right]  ^{2},\nonumber\\
B_{mn}  &  =\sum_{k}\left[  C_{m}^{\left(  k\right)  }\left(  \dfrac
{a^{\left(  k\right)  }}{d_{k}R^{2}}\right)  \left(  C_{n}^{\left(  k\right)
}\right)  ^{\prime}-C_{m}^{\left(  k\right)  }C_{n}^{\left(  k\right)  }%
\dfrac{a^{\left(  k\right)  }}{d_{k}R^{3}}\right]  .\nonumber
\end{align}

When the symmetries given in Table I are used, there can be a considerable
simplification of the expressions for $P_{mn}$ and $Q_{mn}$. For a system of
identical bosons where $a^{\left(  1\right)  }=a^{\left(  2\right)
}=a^{\left(  3\right)  }=a$, $d_{1}=d_{2}=d_{3}=d$ and $C_{n}^{\left(
1\right)  }=C_{n}^{\left(  2\right)  }=C_{n}^{\left(  3\right)  }=C_{n}$,
$P_{mn}$ and $Q_{mn}$ are given by%
\begin{align}
P_{mn}  &  =\dfrac{\sqrt{\varepsilon_{m}^{\prime}\varepsilon_{n}^{\prime}}%
}{\left(  \varepsilon_{m}-\varepsilon_{n}\right)  }\label{BosonPandQ}\\
Q_{mn}  &  =\delta_{mn}\left[  -\dfrac{1}{4}\left(  \dfrac{\varepsilon
_{n}^{\prime\prime}}{\varepsilon_{n}^{\prime}}\right)  ^{2}+\dfrac{1}{6}%
\dfrac{\varepsilon_{n}^{\prime\prime\prime}}{\varepsilon_{n}^{\prime}}\right]
\nonumber\\
&  +\left(  1-\delta_{mn}\right)  \left[  \dfrac{2\varepsilon_{n}^{\prime
}\sqrt{\varepsilon_{n}^{\prime}\varepsilon_{m}^{\prime}}}{\left(
\varepsilon_{m}-\varepsilon_{n}\right)  ^{2}}-\dfrac{\varepsilon_{n}%
^{\prime\prime}}{\left(  \varepsilon_{m}-\varepsilon_{n}\right)  }\sqrt
{\dfrac{\varepsilon_{m}^{\prime}}{\varepsilon_{n}^{\prime}}}\right]
,\nonumber
\end{align}
which are in agreement with previously calculated nonadiabatic corrections for
the three identical boson system \cite{nielsen2001tbp}.

\section{Three distinguishable interacting particles.}

\label{Li6} In this section the adiabatic three-body potentials and the
non-adiabatic couplings are applied to the case of three distinguishable
equal-mass particles. This system has been realized, for instance, in
ultracold three component Fermi gases of $^{6}$Li atoms
\cite{ottenstein2008cst,huckans2009tbr} which has sparked a great deal of
recent theoretical interest
\cite{braaten2009three,naidon2009possible,dincao2009ultracold,rittenhouse2010mfd}%
. The scaling behaviors and recombination rates we discuss in this section can
be found in Ref. \cite{dincao2009ultracold}. We derive them here to illustrate
the power of the methods presented in this paper. The scattering lengths near
the resonance positions used here, as functions of magnetic field, are given
in Refs. \cite{bartenstein2005pdl,ottenstein2008cst,huckans2009tbr} by
\begin{align}
a^{\left(  k\right)  }  &  =a_{b}\left[  1-\dfrac{\Delta}{B-B_{0}}\right]
\left[  1+\alpha\left(  B-B_{0}\right)  \right]  ,\label{scatlength}\\
\text{for }k  &  =1:\nonumber\\
a_{b}  &  =-1450a_{0},\text{ }B_{0}=834.15\text{ G},\text{ }\nonumber\\
\Delta &  =300\text{ G and }\alpha=4\times10^{-4}\text{ G}^{-1};\nonumber\\
\text{for }k  &  =2:\nonumber\\
a_{b}  &  =-1727a_{0},\text{ }B_{0}=690.4\text{ G}\nonumber\\
\text{ }\Delta &  =122.2\text{ G and }\alpha=2\times10^{-4}\text{ G}%
^{-1};\nonumber\\
\text{for }k  &  =3:\nonumber\\
a_{b}  &  =-1490a_{0},\text{ }B_{0}=811.22\text{ G},\text{ }\nonumber\\
\Delta &  =222.3\text{ G and }\alpha=3.95\times10^{-4}\text{ G}^{-1};\nonumber
\end{align}
where $a_{0}$ is the Bohr radius. The Fano-Feshbach resonances in this system
allow for a large variety of tunable interactions.

\begin{figure}[t]
\begin{center}
\includegraphics[width=3.2in]{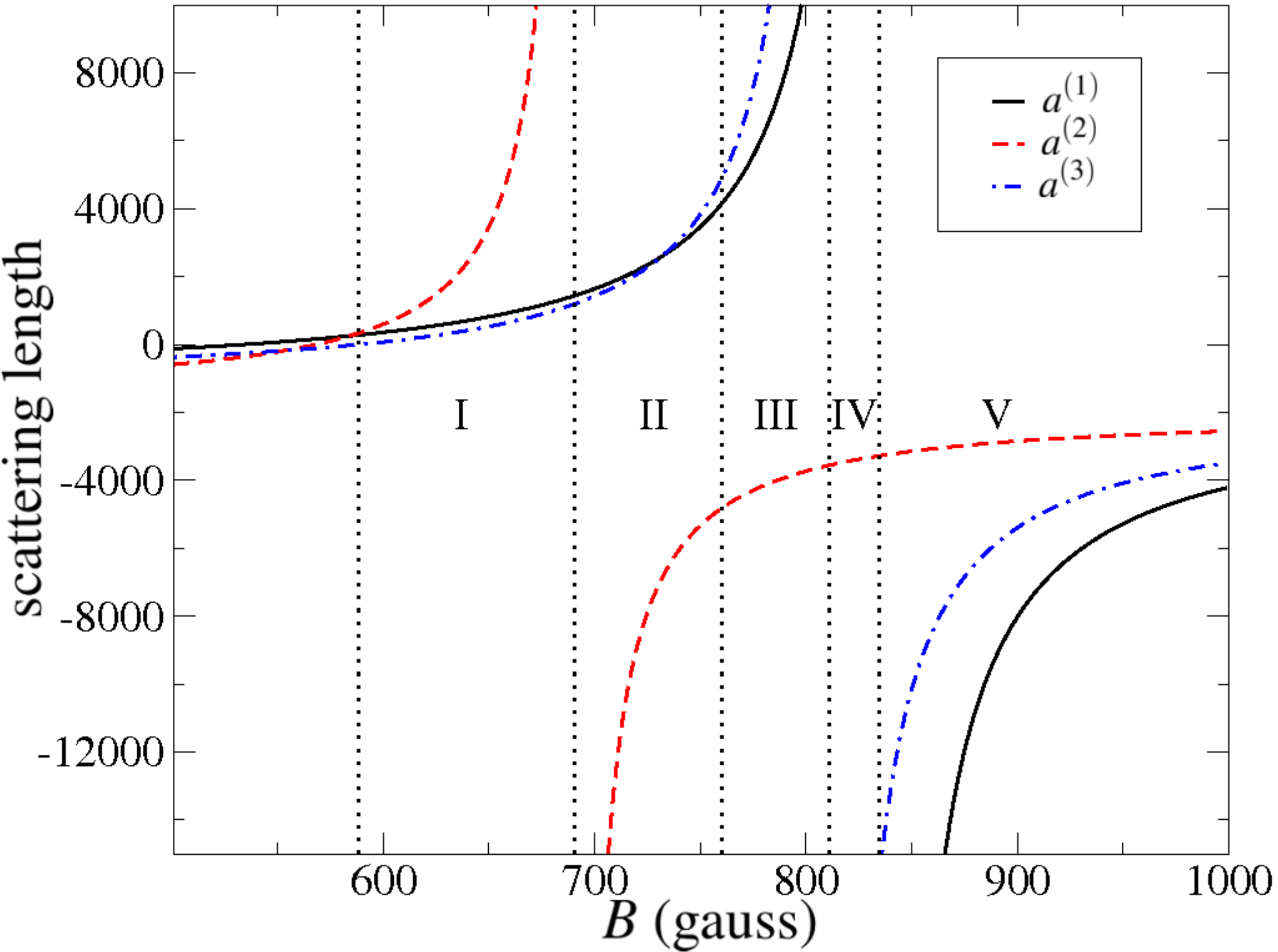}
\end{center}
\caption{(color online) All possible s-wave scattering lengths are shown for
the lowest 3 Zeeman states of Li$^{6}$ from Ref. \cite{bartenstein2005pdl}.
Each marked region gives a different set of length scale discrepancies. Here
$a^{\left(  k\right)  }$ is the scattering length between two atoms in states
$\left\vert i\right\rangle $ and $\left\vert j\right\rangle $ with $k$ as the
component not involved in the interaction.}%
\label{scatlengfig}%
\end{figure}

This series of overlapping resonances produces five different regions of
magnetic field, shown in Fig. \ref{scatlengfig}, near the three resonance
positions, each possessing distinct behavior. In all five regions the
scattering lengths are much larger than the effective range, allowing for the
use of the zero-range interaction assumptions. Table \ref{LengthTab} shows the
various length scale disparities in these regions.

\begin{table}[t]
\begin{center}%
\begin{tabular}
[c]{|c|ll|}\hline
Region &  & \\\hline
I & $r_{0}\ll a^{\left(  3\right)  }\lesssim a^{\left(  1\right)  }\ll
a^{\left(  2\right)  }$ & \multicolumn{1}{|l|}{$a^{\left(  1\right)
},a^{\left(  2\right)  },a^{\left(  3\right)  }>0$}\\\hline
II & $r_{0}\ll a^{\left(  3\right)  }\sim a^{\left(  1\right)  }\ll\left\vert
a^{\left(  2\right)  }\right\vert $ & \multicolumn{1}{|l|}{$a^{\left(
2\right)  }<0;a^{\left(  1\right)  },a^{\left(  3\right)  }>0$}\\\hline
III & $r_{0}\ll\left\vert a^{\left(  2\right)  }\right\vert \ll a^{\left(
1\right)  },a^{\left(  3\right)  }$ & \multicolumn{1}{|l|}{$a^{\left(
2\right)  }<0;a^{\left(  1\right)  },a^{\left(  3\right)  }>0$}\\\hline
IV & $r_{0}\ll\left\vert a^{\left(  2\right)  }\right\vert \ll\left\vert
a^{\left(  1\right)  }\right\vert ,a^{\left(  3\right)  }$ &
\multicolumn{1}{|l|}{$a^{\left(  2\right)  },a^{\left(  1\right)
}<0;a^{\left(  3\right)  }>0$}\\\hline
V & $r_{0}\ll\left\vert a^{\left(  2\right)  }\right\vert \ll\left\vert
a^{\left(  1\right)  }\right\vert \ll\left\vert a^{\left(  3\right)
}\right\vert $ & \multicolumn{1}{|l|}{$a^{\left(  1\right)  },a^{\left(
2\right)  },a^{\left(  3\right)  }<0$}\\\hline
\end{tabular}
\end{center}
\caption{The possible tunable interaction regimes near the resonances of
$^{6}$Li are given.}%
\label{LengthTab}%
\end{table}

\begin{figure}[th]
\begin{center}
\includegraphics[width=3.2in]{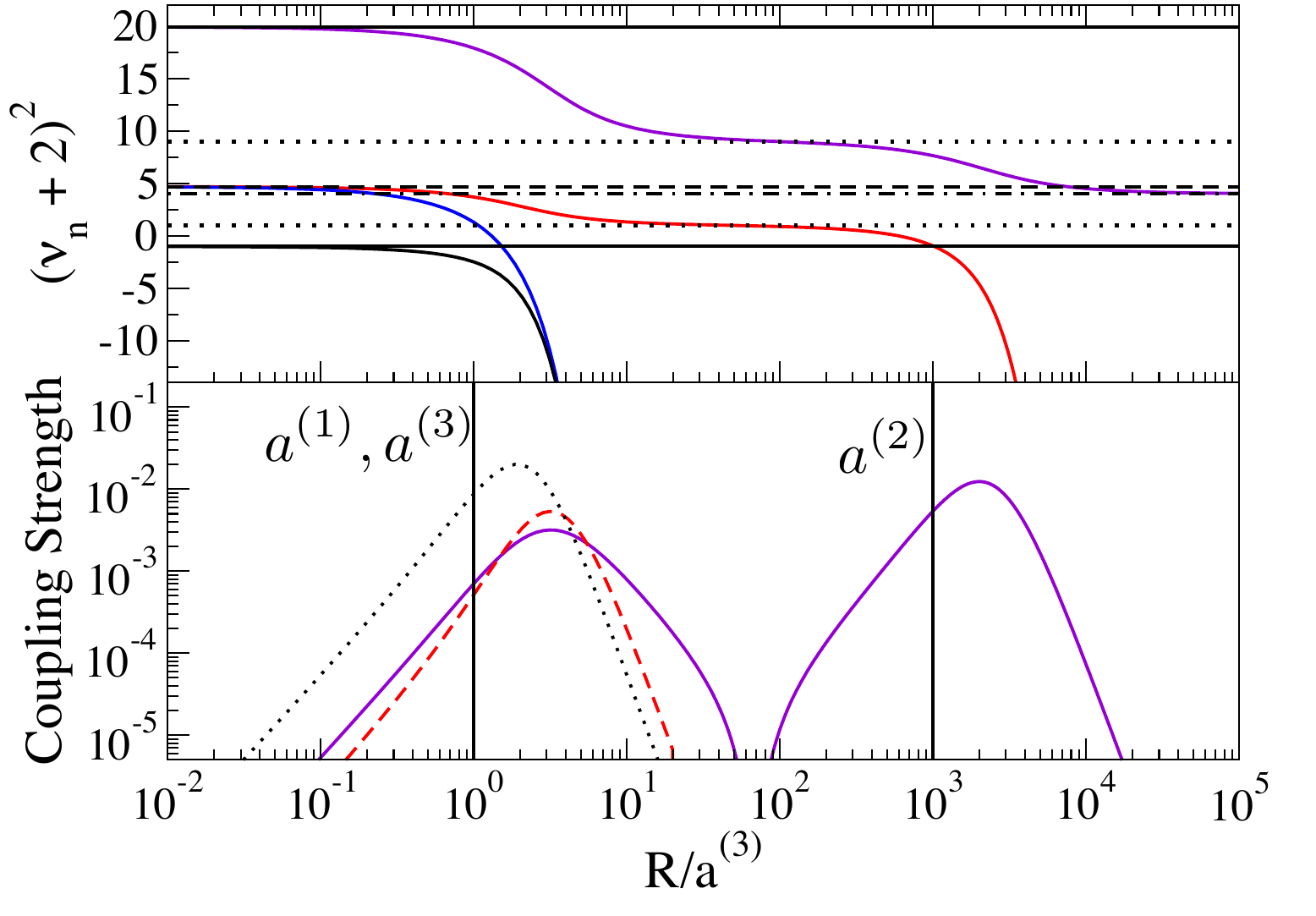}
\end{center}
\caption{(color online) (a) For an example system having $a^{(1)}=a^{(3)}$ and
$a^{(2)}=1000a^{(1)}$, the first four hyperangular eigenvalues are shown as
functions of the hyperradius. The solid black horizontal lines show the
expected behavior for 3 identical resonantly interacting bosons. The dashed
line gives the behavior of two identical fermions interacting resonantly with
a third distinguishable particle. Dotted lines give the expected universal
behavior for a single resonant scattering length. Finally, the dot-dashed line
is the lowest expected free space eigenvalue for three distinguishable free
particles.(b)The coupling strengths between the third and fourth (purple solid
curve), the first and fourth (red dashed curve), and the first and third
(black dotted curve) adiabatic potentials are shown as a function of $R$.}%
\label{Hypangvalsfig}%
\end{figure}

Figure \ref{Hypangvalsfig}(a) shows an example of the lowest four hyperangular
eigenvalues $\left(  \nu+2\right)  ^{2}$ obtained from solving
Eq.~(\ref{TrancendME}) for $a^{\left(  1\right)  }=a^{\left(  3\right)  }$ and
$a^{\left(  2\right)  }=1000a^{\left(  1\right)  }$. This is provided as an
example that is qualitatively similar to the behavior of the system in region
I. When the hyperradius is in a region where all other length scales are much
different, the hyperangular eigenvalue $\left(  \nu+2\right)  ^{2}$ becomes
constant, or, in the case of 2-body bound states, becomes proportional to
$R^{2}$. This behavior can be interpreted as giving a universal set of
potential curves from Eq.~(\ref{EffH}). For example in region I where
$r_{0}\ll a^{\left(  3\right)  }\lesssim a^{\left(  1\right)  }\ll a^{\left(
2\right)  }$ there are three hyperradial regions: $r_{0}\ll R\ll a^{\left(
3\right)  }\lesssim a^{\left(  1\right)  }\ll a^{\left(  2\right)  }$;
$r_{0}\ll a^{\left(  3\right)  }\lesssim a^{\left(  1\right)  }\ll R\ll
a^{\left(  2\right)  }$; and $r_{0}\ll a^{\left(  3\right)  }\lesssim
a^{\left(  1\right)  }\ll a^{\left(  2\right)  }\ll R$. In each region the
hyperangular eigenvalues take on the universal value that is expected for
resonant interactions \cite{efimov1973,nielsen2001tbp,dincao2005sls}.

\begin{figure}[th]
\begin{center}
\includegraphics[width=3.5in]{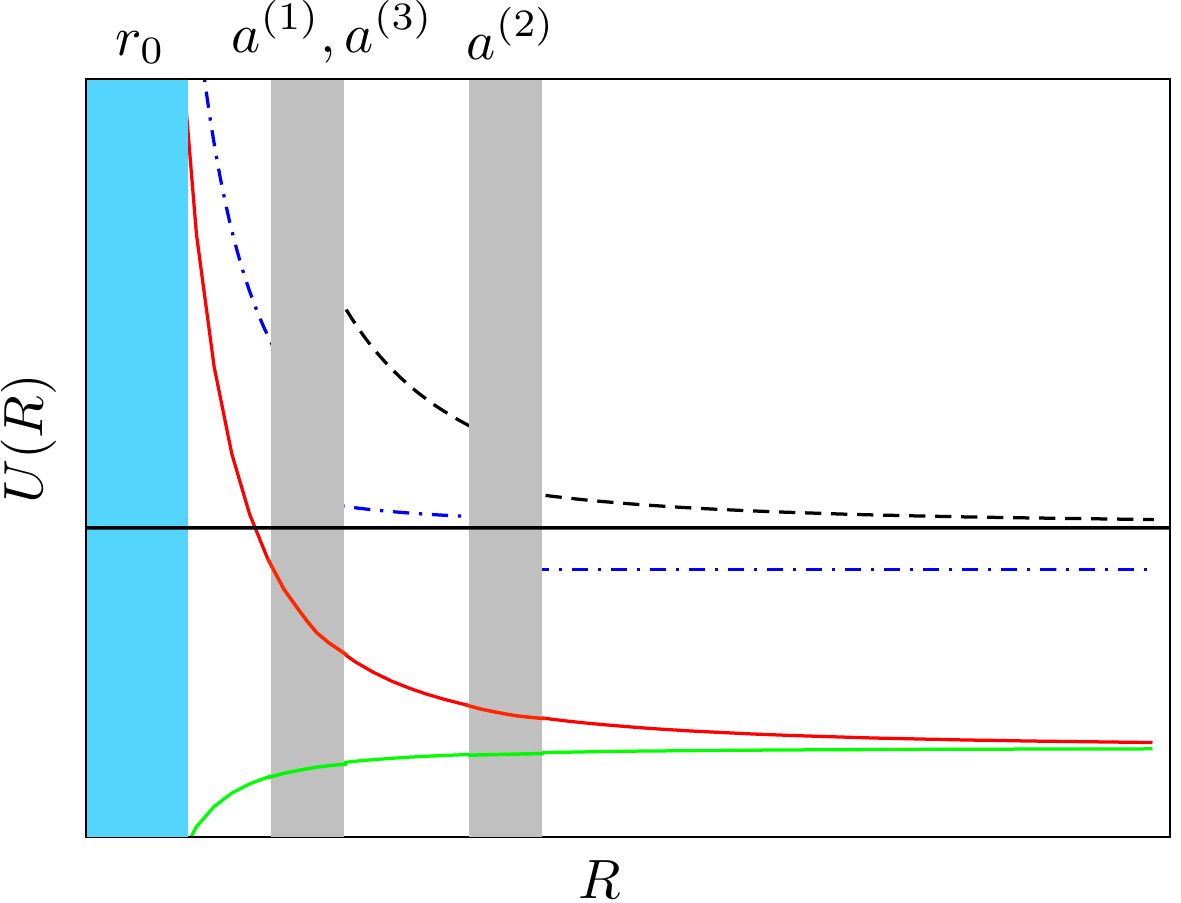}
\end{center}
\caption{(color online) A schematic picture of the first four hyperradial
potentials in region I is shown. The grey areas, labeled by the appropriate
scattering lengths, indicate regions where the potentials are changing from
one universal behavior to another. The blue region, labeled by $r_{0}$,
indicates the short range region where the zero-range pseudo-potential not
longer can be applied.}%
\label{PotSchemR1}%
\end{figure}

Figure \ref{PotSchemR1} schematically shows the behavior of the first few
hyperradial effective potentials from Eq. (\ref{EffH}). The grey areas are the
regions where potentials are transitioning from one universal behavior to the
next. The zero-range pseudo-potential cannot describe the short range details
of the interaction, meaning that the potentials found here are only valid for
$R\gg r_{0}$ where $r_{0}$ is a short range parameter shown schematically as
the labeled blue region of Fig. \ref{PotSchemR1}. 

\begin{figure}[th]
\begin{center}
\includegraphics[width=3.5in]{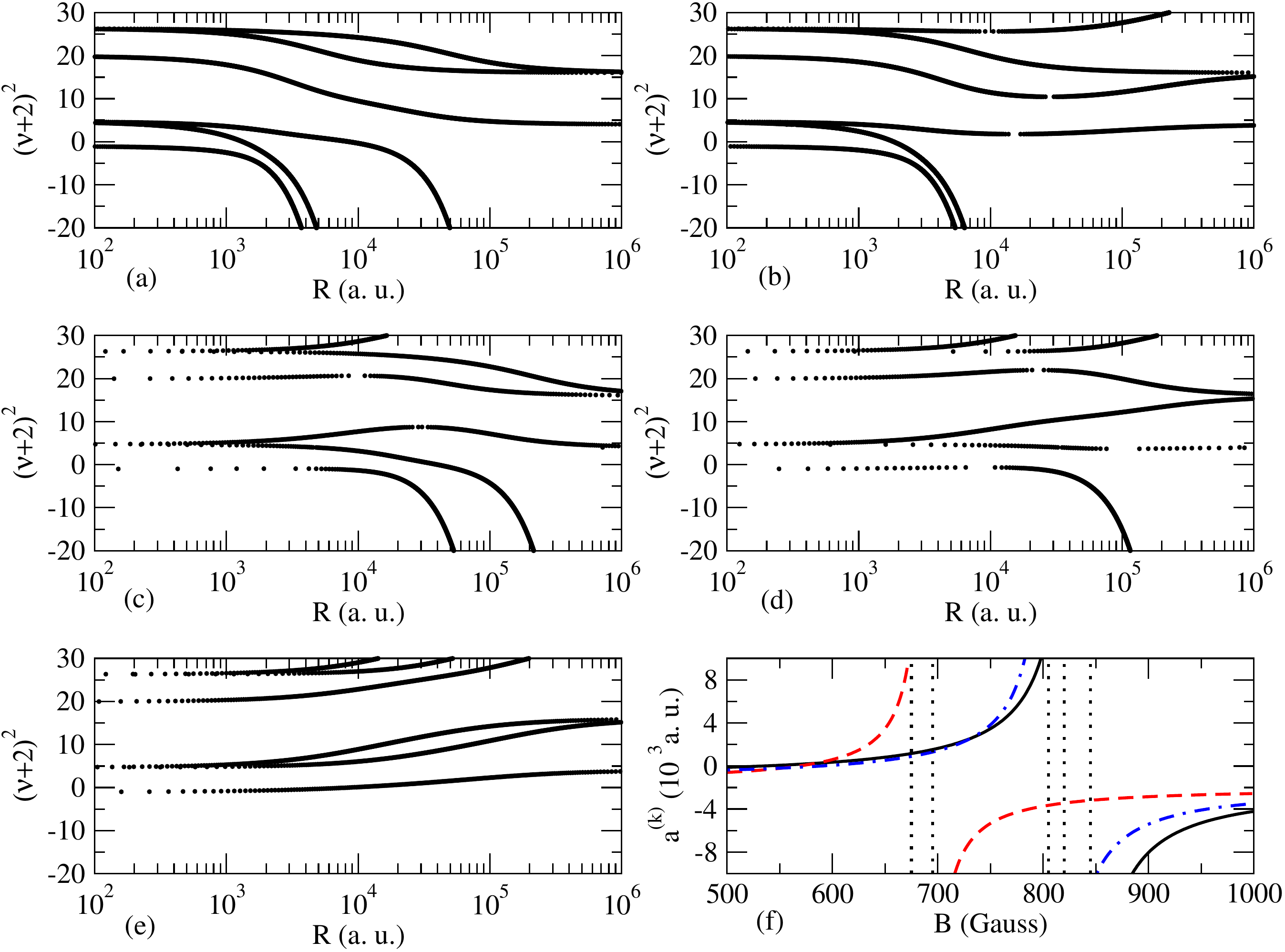}
\end{center}
\caption{(color online) (a)-(e) Examples of the hyperangular eigenvalues from
each region of magnetic field are shown as a function of the hyperradius in
atomic units. (f) The three s-wave scattering lengths are shown as a reference
plotted versus the magnetic field strength. The vertical dotted lines, from
right to left, show the magnetic fields at which the hyperangular eigenvalues
in (a)-(e) were found, $B=675,695,805,820$ and $845$ gauss respectively.}%
\label{Hypangvalsfig2}%
\end{figure}

Figure \ref{Hypangvalsfig}(b) shows the coupling strength, $P_{mn}^{2}%
/2\mu\left[  u_{m}\left(  R\right)  -u_{n}\left(  R\right)  \right]  $,
between the different potentials. The places where this coupling peaks are the
points where a transition between curves is the most probable. Figures
\ref{Hypangvalsfig2}(a-e) are examples of the hyperangular eigenvalues found
in each region. The magnetic field at which each set of eigenvalues are found
is shown as dotted lines in Fig. \ref{Hypangvalsfig2}(f) from left to right
for Fig. \ref{Hypangvalsfig2}(a-e) respectively. In each figure the
hyperangular eigenvalue can be seen flattening out to a universal constant in
each region of length scale discrepancy. As the magnetic field is scanned
through each resonance, one two-body bound state becomes a virtual state. This
behavior can be seen in the hyperangular eigenvalues that diverge toward
$-\infty$. As each resonance is crossed, one of the hyperangular eigenvalue
curves goes from diverging to $-\infty$ to converging to $\left(
\nu+2\right)  ^{2}\rightarrow4$. 

As a final examination of this system, we extract the scaling of the low
energy three body recombination rate, i.e. the rate at which three particles
collide and form a dimer and a free particle. The lowest 3-body curve, the
lowest potential that goes to the three-free-particle threshold, is the
potential that dominates this process. Contributions from higher hyperradial
potentials will be suppressed due to larger tunneling barriers. One limitation
of the zero-range pseudo-potential is that it only admits at most one dimer of
each type. The process of three-body recombination releases the binding energy
of the dimer state as kinetic energy between the dimer and remaining particle.
For the purposes of this study we will concentrate on the three-body
recombination processes that result in trap loss processes, where the energy
released in the recombination can be assumed sufficient to eject the remaining
fragments from a trap.

The event rate coefficient for $N$ initially unbound particles with total
orbital angular momentum $L$ to make a transition from a hyperspherical
potential curve with hyperangular eigenvalue $\lambda$ to a lower lying final
state is given by \cite{mehta2009gtd,esry1999rta}%
\begin{equation}
K_{N}=\dfrac{\hbar k}{\mu}N_{S}\left(  \dfrac{2\pi}{k}\right)  ^{d-1}%
\dfrac{\Gamma\left(  \dfrac{d}{2}\right)  }{2\pi^{d/2}}\sum_{i,f}\left(
2L+1\right)  \left\vert T_{fi}\right\vert ^{2}, \label{Eq:Nbodrecomb}%
\end{equation}
where $d$ is the total dimension of the system (in the case of three-body
recombination $d=6$), $T_{fi}\equiv S_{fi}-\delta_{fi}$ is the transition
matrix element between an initial three-body entrance channel $i$ and a final
exit channel $f$, and $k=\sqrt{2\mu E}/\hbar$ is the wave number of the
asymptotic hyperradial wavefunction. The sum in this equation runs over all
the initial, asymptotic channels with total angular momentum $L$ that
contribute to the scattering process. In Eq.~(\ref{Eq:Nbodrecomb}), $N_{S}$ is
the number of permutational symmetries in the system. For three
distinguishable particles $N_{S}=1$, but it can be different, for instance for
$N$ identical bosons, $N_{S}=N!$.

In the low energy regime, only the lowest $L=0$ initial three-body channel
will contribute, while higher channels will be suppressed. The sum over final
$T$ matrix elements can be approximated using the Wentzel--Kramers--Brillouin
(WKB) phase in the entrance channel \cite{mehta2009gtd,dincao2005sls}:%
\begin{equation}
\sum_{f}\left\vert T_{fi}\right\vert ^{2}\approx\dfrac{e^{-2\gamma}}{2}%
\dfrac{\sinh2\eta}{\cos^{2}\phi+\sinh^{2}\eta}, \label{Eq:WKBtunn}%
\end{equation}
where $\eta$ is an imaginary phase which parameterizes the losses from the
incoming channel. In Eq.~(\ref{Eq:WKBtunn}) $\gamma$ is the total WKB
tunneling integral between the outer classical turning point and the
hyperradial position at which the transition to the outgoing state occurs,
i.e.%
\begin{equation}
\gamma=\operatorname{Re}\left[  \int_{R_{0}}^{R_{T}}\sqrt{\dfrac{2\mu}%
{\hbar^{2}}\left[  U\left(  R\right)  -E\right]  +\dfrac{1}{4R^{2}}%
}\;dR\right]  , \label{Eq:WKBtunneling}%
\end{equation}
where $E$ is the initial three body energy, $R_{T}$ is the outer classical
turning point and $R_{0}$ is the position at which the coupling between the
incoming and outgoing channels peaks. In Eq.~(\ref{Eq:WKBtunn}) $\phi$ is the
WKB phase acumulated in any inner attractive well:%
\begin{equation}
\phi=\operatorname{Im}\left[  \int_{R_{0}}^{R_{T}}\sqrt{\dfrac{2\mu}{\hbar
^{2}}\left[  U\left(  R\right)  -E\right]  +\dfrac{1}{4R^{2}}}\;dR\right]  .
\label{EQ:WKBphase}%
\end{equation}
The extra repulsive $1/4R^{2}$ term in Eqs.~(\ref{Eq:WKBtunneling}) and
(\ref{EQ:WKBphase}) is due to the Langer correction \cite{langer1937cfa}. The
total $T$-matrix element will depend on the detailed nature of the real short
range interactions and the behavior of the outgoing channels, but the scaling
behavior with the scattering lengths will be determined by
Eq.~(\ref{Eq:WKBtunn}). In each region of magnetic field, there are different
length scale discrepancies and different numbers of bound states. As a result,
we will examine each region separately.

\subsubsection{Region I $(a^{\left(  1\right)  }\sim a^{\left(  3\right)  }\ll
a^{\left(  2\right)  })$}

Figure~\ref{Hypangvalsfig2}(a) shows the behavior of the first few
hyperangular eigenvalues in region I. The first three eigenvalues correspond
to dimer states, while the fourth corresponds to the lowest three-body
potential and is the entrance channel that will control three-body
recombination. The lower two dimer states are relatively deeply bound with
binding energies, $\hbar^{2}/ma^{2}$, on the order of $10^{-12}$ Hartree. This
is comparable to the trap depth energy of a normal magneto-optical trap for
experiments with $^{6}$Li \cite{ottenstein2008cst,huckans2009tbr}, meaning
that recombination into these dimer channels typically releases enough energy
to eject the remaining dimer-atom system from the trap.

In the limit where $R\gg a^{\left(  3\right)  }$, the three atoms are far
enough apart to be in the non-interacting regime. This means that the
hyperangular eigenfunction limits to the lowest allowed three-body
hyperspherical harmonic with its corresponding eigenvalue, $\left(
\nu+2\right)  ^{2}\rightarrow4.$ In this limit the hyperradial potential
becomes%
\begin{equation}
U\left(  R\gg a^{\left(  3\right)  }\right)  =\dfrac{\hbar^{2}}{2\mu}%
\dfrac{4-1/4}{R^{2}}.
\end{equation}
For very low energy scattering, the classical turning point in
Eq.~(\ref{Eq:WKBtunn}), is approximately%
\begin{equation}
R_{T}=\dfrac{1}{2k}. \label{Eq:classicaltp}%
\end{equation}
In fact, this will be the turning point for all of the three-body
recombination processes discussed in this section.

\begin{figure}[th]
\begin{center}
\includegraphics[width=3.5in]{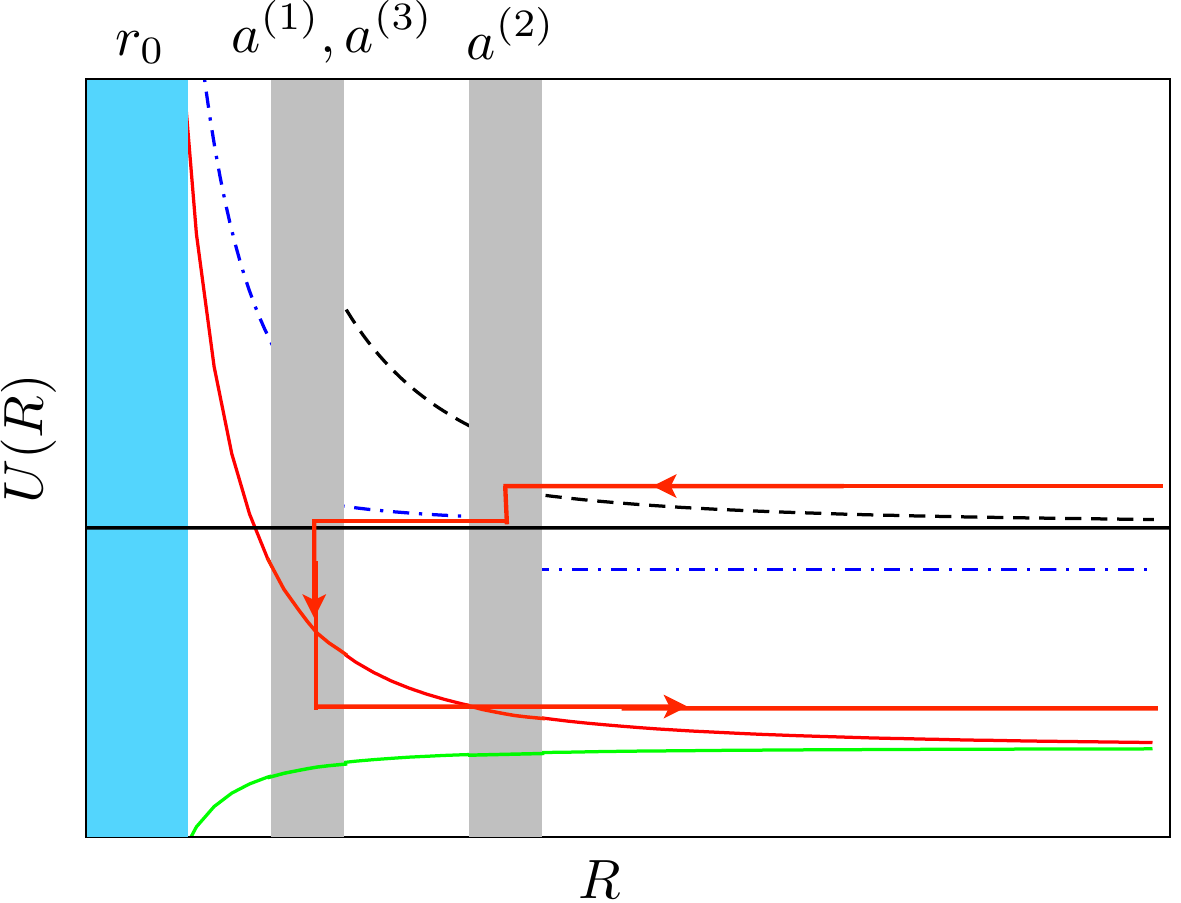}
\end{center}
\caption{(Color online) A schematic of the path for three-body recombination
in region I is shown. Transition regions are labeled by the appropriate length
scale, and the short-range non-universal region is labeled by $r_{0}$.}%
\label{PathR1}%
\end{figure}

It is possible for recombination to occur directly between the lowest
three-body curve and the deep dimer channels, but this direct process is
strongly suppressed due to the large tunneling barrier in the three-body
potential at small $R$. The favored path is through a transition to the weakly
bound dimer channel, shown schematically in Fig. \ref{PathR1}.
The coupling between the lowest three-body channel and the weakly
bound dimer channel peaks at approximately $R\sim a^{\left(  2\right)  }$,
while the coupling peak between the weakly bound dimer channel and the
remaining two dimer channels occurs at approximately $R\sim a^{\left(
3\right)  }$ $\sim a^{\left(  1\right)  }$. In the regime where $a^{\left(
1\right)  }\sim a^{\left(  3\right)  }\ll R\ll a^{\left(  2\right)  }$ the
three particles are so far apart that they cannot see the smaller scattering
lengths $a^{\left(  1\right)  }$ and $a^{\left(  3\right)  }$, but the third
scattering length is so large compared to the hyperradius that it might as
well be infinite. This leads to a universal potential whose hyperangular
eigenvalue can be found by solving Eq.~(\ref{TrancendME}) with $a^{\left(
1\right)  }=a^{\left(  3\right)  }=0$ and $a^{\left(  2\right)  }%
\rightarrow\infty$, i.e%
\begin{align}
U\left(  R\right)   &  =\dfrac{\hbar^{2}}{2\mu}\dfrac{p_{1}^{2}-1/4}{R^{2}%
},\label{Eq:Univpot1res}\\
p_{1}  &  =\left(  \nu+2\right)  =1.\nonumber
\end{align}
This intermediate universal behavior can clearly be seen in Fig.
(\ref{Hypangvalsfig}a).

The behavior of each channel can be approximated by the universal behavior of
the hyperradial potential in each region. Under this assumption, using
Eq.~(\ref{Eq:WKBtunn}), the tunneling probability is given by%
\begin{align}
e^{-2\gamma}  & \propto\exp\left[  -2\int_{a^{\left(  3\right)  }}^{a^{\left(
2\right)  }}dR\sqrt{\dfrac{p_{1}^{2}}{R^{2}}-\frac{2\mu}{\hbar^{2}}E}\right.
\\
& -\left.  2\int_{a^{\left(  2\right)  }}^{R_{T}}dR\sqrt{\dfrac{4}{R^{2}%
}-\frac{2\mu}{\hbar^{2}}E}\right]  .\nonumber
\end{align}
If the scattering energy is very small, $E\ll\hbar^{2}/m\left[  a^{\left(
2\right)  }\right]  ^{2}$, then the energy dependence in these integrands
becomes negligible leaving,
\begin{equation}
e^{-2\gamma}\propto k^{4}\left(  a^{\left(  2\right)  }a^{\left(  3\right)
}\right)  ^{2}.\label{Eq:Scaling1}%
\end{equation}
Inserting this in for the $T$-matrix element in Eq.~(\ref{Eq:Nbodrecomb})
gives the scaling behavior of the recombination rate with the scattering
lengths \cite{dincao2005sls}:%
\begin{equation}
K_{3}\propto\left(  a^{\left(  2\right)  }a^{\left(  3\right)  }\right)
^{2}.\label{Eq:RecomrateR1}%
\end{equation}
It was assumed here the final transition occurs at $R\sim a^{\left(  3\right)
}$ leading to the scaling behavior with $a^{\left(  3\right)  }$, but the
transition could just as easily have occurred at $R\sim a^{\left(  1\right)
}$. $a^{\left(  1\right)  }$ and $a^{\left(  3\right)  }$ are approximately
equal here, and which one dominates the transition depends on the short range
behavior of the real two-body interaction. To extract the scaling behavior
with respect to $a^{\left(  1\right)  }$, one can simply replace $a^{\left(
3\right)  }$ with $a^{\left(  1\right)  }$ in Eq.~(\ref{Eq:RecomrateR1}) as
long as $a^{\left(  1\right)  }$ and $a^{\left(  3\right)  }$ are
approximately equal.

\subsubsection{Region II $(a^{\left(  1\right)  }\sim a^{\left(  3\right)
}\ll\left\vert a^{\left(  2\right)  }\right\vert )$}

The recombination in region II is simpler than in region I as there is no
weakly bound intermediate state. Again, we assume that the trap loss
recombination is dominated by transitions to the two remaining dimer states
seen in Fig. \ref{Hypangvalsfig2}(b). The lowest three-body potential has
coupling to these channels that peaks at $R\sim a^{\left(  1\right)  }$ and
$R\sim a^{\left(  3\right)  }$. For $R\gg\left\vert a^{\left(  2\right)
}\right\vert $ the hyperangular eigenvalue takes on the non-interacting value
$\left(  \nu+2\right)  \rightarrow2$. For $a^{\left(  1\right)  },a^{\left(
3\right)  }\ll R\ll\left\vert a^{\left(  2\right)  }\right\vert $ the
universal hyperangular eigenvalue $\left(  \nu+2\right)  =p_{1}=1$ is seen
again \cite{nielsen2001tbp,dincao2005sls,Braaten2006physrep}. Ignoring the
transitional region between these two regimes the transition probability is
given by
\begin{equation}
e^{-2\gamma}\propto\exp\left[  -2\left(  \int_{a^{\left(  3\right)  }%
}^{\left\vert a^{\left(  2\right)  }\right\vert }dR\sqrt{\dfrac{p_{1}^{2}%
}{R^{2}}}+\int_{\left\vert a^{\left(  2\right)  }\right\vert }^{R_{T}}%
dR\sqrt{\dfrac{4}{R^{2}}}\right)  \right]  .
\end{equation}
Inserting this into Eq.~(\ref{Eq:Nbodrecomb}) gives a recombination rate that
has the same scaling behavior as in region I \cite{dincao2005sls}:%
\begin{equation}
K_{3}\propto\left(  a^{\left(  2\right)  }a^{\left(  3\right)  }\right)  ^{2}.
\label{Eq:RecomrateR2}%
\end{equation}
Again it is assumed that the final transition occurs at $R\sim a^{\left(
3\right)  }$, but it could occur at $a^{\left(  1\right)  }$ as well. As in
Region I, the scaling behavior with respect to $a^{\left(  1\right)  }$ can be
found by simply replacing $a^{\left(  3\right)  }$ with $a^{\left(  1\right)
}$ in Eq.~(\ref{Eq:RecomrateR1}) as long as $a^{\left(  1\right)  }$ and
$a^{\left(  3\right)  }$ are close.

\subsubsection{Region III $(\left\vert a^{\left(  2\right)  }\right\vert \ll
a^{\left(  1\right)  }\sim a^{\left(  3\right)  })$ and Region IV $(\left\vert
a^{\left(  2\right)  }\right\vert \ll\left\vert a^{\left(  1\right)
}\right\vert \sim a^{\left(  3\right)  })$}

In Region III, none of the dimers predicted by the zero-range model have
enough binding energy to cause trap loss. While recombination can occur into
these channels, we will focus on the process of recombination to deeply bound
states here. In reality, the deep interaction potential between two Li atom in
different spin states admits many deeply bound dimer states, and a true
hyperspherical description of the system would have channels going to each
possible dimer-atom threshold. The energy released in recombining into these
deep states is enough to kick the atoms out of any normal trap. Because the
deeply bound states are of the size of the range of the interaction, coupling
to the deeply bound hyperradial channels will peak at small hyperradius,
$R\sim r_{0}$, and the rate can be found by studying the tunneling probability
of reaching these states.

As with the recombination process in Region I, the most favorable pathway
involves multiple steps. Starting from the lowest three-body channel, a
transition is made to either the first or second weakly bound dimer channel.
Because $a^{\left(  1\right)  }$ and $a^{\left(  3\right)  }$ are similar in
magnitude, the coupling to these channels peaks in the same region. If the
transition is made to the highest dimer channel, then another transition is
made directly to the second.

\begin{figure}[th]
\begin{center}
\includegraphics[width=3.5in]{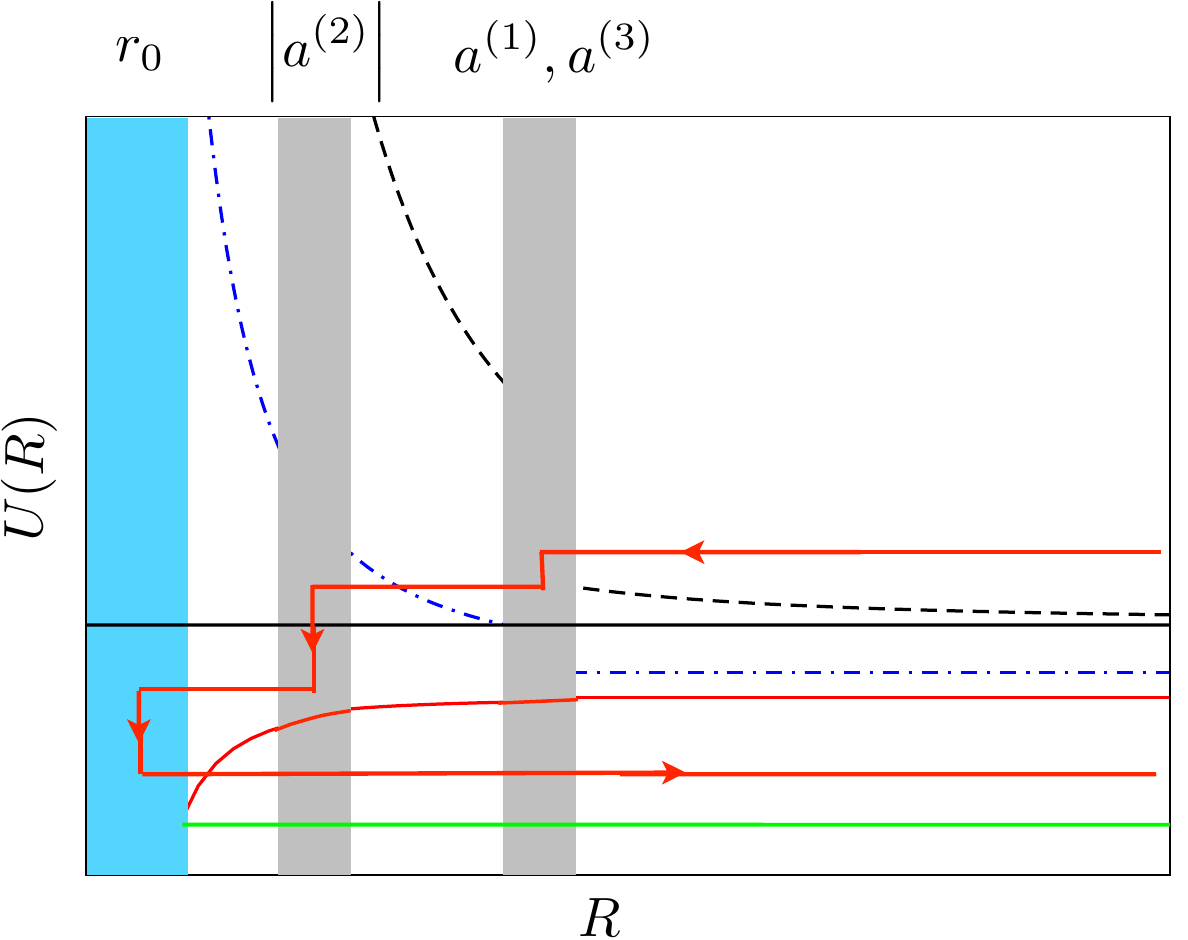}
\end{center}
\caption{(color online) A schematic of the potentials and the path for
three-body recombination in region III is shown. Again the labeled grey
regions indicate a transition from one universal potential behavior to
another. The green line represents the hyperradial potential for a deeply
bound dimer state. The blue area labeled by $r_{0}$ is the short range region
not described by zero-range interactions.}%
\label{PathSchem3}%
\end{figure}

This pathway is shown schematically in Fig. \ref{PathSchem3}.
An interesting thing occurs in the lowest weakly bound potential
when $\left\vert a^{\left(  2\right)  }\right\vert \ll R\ll a^{\left(
1\right)  }\sim a^{\left(  3\right)  }$: the universal potential becomes
attractive. This region of attractive potential gives rise to a number of
phenomena. For instance, in the limit $a^{\left(  1\right)  },a^{\left(
3\right)  }\rightarrow\infty$, the universal attractive potential supports an
infinite number of geometrically spaced three-body bound states, giving rise
to the Efimov effect. In the process of three-body recombination to
deeply-bound dimer states, though, there is no tunnelling suppression in this
channel, and the hyperradial wavefunction merely accumulates phase in this
region. As a result the WKB tunnelling probability is controlled by the
transition at $R\sim a^{\left(  1\right)  },a^{\left(  3\right)  }$:
\begin{equation}
e^{-2\gamma}\propto\exp\left[  -2\int_{a^{\left(  1\right)  }}^{R_{T}}%
dR\sqrt{\dfrac{4}{R^{2}}}\right]  .
\end{equation}
Inserting this into Eq.~(\ref{Eq:Nbodrecomb}) gives the scaling of three-body
recombination to deep dimer states as%
\begin{equation}
K_{3}\propto\left[  a^{\left(  1\right)  }\right]  ^{4}.
\label{Eq:RecomrateR3}%
\end{equation}
Again, it is assumed here that $a^{\left(  1\right)  }$ and $a^{\left(
3\right)  }$ are similar in magnitude. If this is not the case, for instance
if $a^{\left(  1\right)  }\gg a^{\left(  3\right)  }$, then a scaling behavior
similar to that of Eq.~(\ref{Eq:RecomrateR1}) is recovered:%
\begin{equation}
K_{3}\propto\left[  a^{\left(  1\right)  }a^{\left(  3\right)  }\right]  ^{2}.
\label{Eq:RecomrateR3c2}%
\end{equation}

In Region IV there is only a single weekly bound dimer state available, and
trap loss will occur through recombination to deeply bound dimers. The path
here is similar to that of Region III, where a transition happens from the
lowest three-body channel to the weakly bound dimer channel. From there the
hyperradial wavefunction can go to the small $R$ region without further
suppression. This process then yields the same three-body recombination
scaling behavior as Eq.~(\ref{Eq:RecomrateR3}) when $a^{\left(  1\right)
}\sim\left\vert a^{\left(  3\right)  }\right\vert $. When $a^{\left(
3\right)  }\gg\left\vert a^{\left(  1\right)  }\right\vert ,$ the scaling
predicted by Eq.~(\ref{Eq:RecomrateR3c2}) is recovered.

\subsubsection{Region V $(\left\vert a^{\left(  2\right)  }\right\vert
\ll\left\vert a^{\left(  3\right)  }\right\vert \ll\left\vert a^{\left(
1\right)  }\right\vert )$}

\begin{figure}[th]
\begin{center}
\includegraphics[width=3.5in]{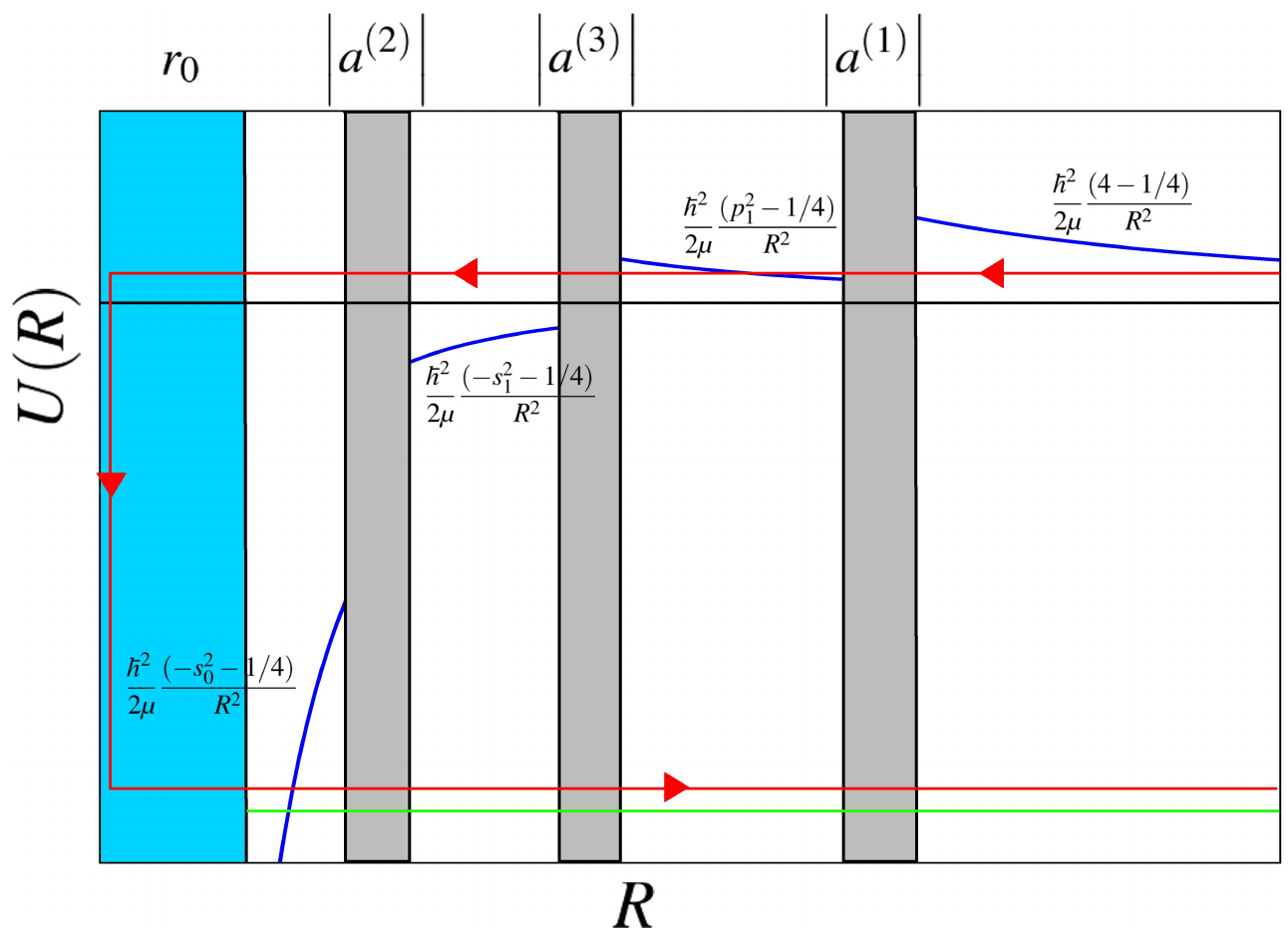}
\end{center}
\caption{(Color online) A schematic of the lowest hyperradial potential is
shown with the path for three-body recombination to deeply bound states. The
green line represents the hyperradial potential for a deeply bound dimer
state. Labeled grey areas indicate transition regions from one universal
behavior to another, and the blue region indicates the short range regime.}%
\label{PathR5}%
\end{figure}

In this regime the recombination process is entirely controlled by the lowest
three-body channel, shown schematically in Fig. \ref{PathR5}.
The hyperradial potential has three universal regimes. The first,
when $r_{0}\ll R\ll\left\vert a^{\left(  2\right)  }\right\vert \ll\left\vert
a^{\left(  3\right)  }\right\vert \ll\left\vert a^{\left(  1\right)
}\right\vert $, is identical to that of three strongly interacting bosons. The
hyperangular eigenvalue, $\left(  \nu+2\right)  ^{2}$, is the first solution
to Eq.~(\ref{BBBeq}) in the limit where $R/a\rightarrow0$, yielding the
hyperradial potential,%
\begin{align}
U\left(  R\right)   &  =\dfrac{\hbar^{2}}{2\mu}\dfrac{-\left(  s_{0}\right)
^{2}-1/4}{R^{2}},\\
s_{0}  &  =1.0062.\nonumber
\end{align}
In the next regime, when $r_{0}\ll\left\vert a^{\left(  2\right)  }\right\vert
\ll R\ll\left\vert a^{\left(  3\right)  }\right\vert \ll\left\vert a^{\left(
1\right)  }\right\vert $, the three particles are far enough apart so as not
to see the smallest scattering length. As a result the hyperangular eigenvalue
is governed by Eq.~(\ref{TrancendME}) with the BBX symmetry of Table
\ref{Symmtab} imposed:%
\begin{align}
U\left(  R\right)   &  =\dfrac{\hbar^{2}}{2\mu}\dfrac{-\left(  s_{1}\right)
^{2}-1/4}{R^{2}},\label{Eq:2respot}\\
s_{1}  &  =0.4137.\nonumber
\end{align}
In the regime where $r_{0}\ll\left\vert a^{\left(  2\right)  }\right\vert
\ll\left\vert a^{\left(  3\right)  }\right\vert \ll R\ll\left\vert a^{\left(
1\right)  }\right\vert $, there is only one scattering length that is seen by
the system, and the universal potential becomes that of
Eq.~(\ref{Eq:Univpot1res}). In the final regime, where the hyperradius is much
larger than all of the scattering lengths, the potential goes to the
non-interacting behavior of a hyperspherical harmonic.

The transition to a deeply bound dimer state occurs at $R\sim r_{0}$ following
the path shown in Fig. \ref{PathR5}. To get to this region, the wavefunction
must first tunnel through a barrier, leading to suppression of the
recombination rate. Once through the barrier, the wavefunction accumulates
phase in the attractive potential regime. If enough phase can be accumulated
in this regime, then a three-body bound state (a so called Efimov state) can
be present leading to a resonance in the recombination rate. The final
recombination rate for this process is
\cite{esry1999rta,nielsen2001tbp,dincao2005sls,braaten2004edr}
\begin{equation}
K_{3}\propto A\dfrac{\sinh2\eta}{\cos^{2}\left(  \phi_{WKB}\right)  +\sinh
^{2}\eta}, \label{Eq:RecomrateR5Efi}%
\end{equation}
where $\eta$ is controlled by the short range properties of the system,
$\phi_{WKB}$ is the WKB phase accumulated in the attractive regime from
$r_{0}$ to $\left\vert a^{\left(  3\right)  }\right\vert $, and $A$ is
proportional to the tunneling suppression through the barrier:%
\begin{align}
A  &  \propto\left[  a^{\left(  3\right)  }a^{\left(  1\right)  }\right]
^{2},\label{Eq:ScalingR5}\\
\phi_{WKB}  &  =s_{1}\ln\left(  \dfrac{a^{\left(  3\right)  }}{a^{\left(
2\right)  }}\right)  +s_{0}\ln\left(  \dfrac{\left\vert a^{\left(  2\right)
}\right\vert }{r_{0}}\right)  .
\end{align}

\begin{figure}[th]
\begin{center}
\includegraphics[width=3.5in]{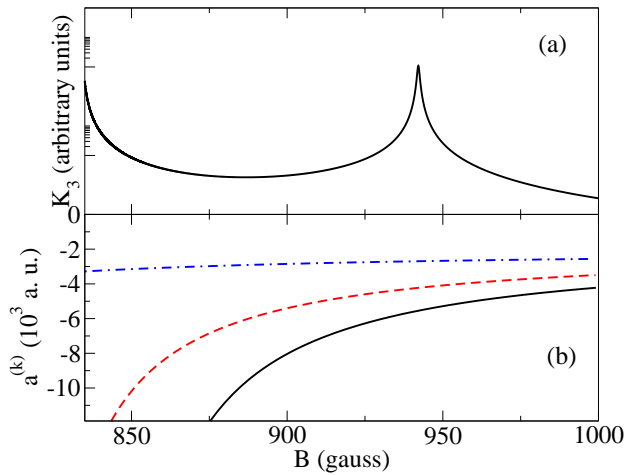}
\end{center}
\caption{(color online) (a)The three-body recombination rate from
Eq.~(\ref{Eq:RecomrateR5Efi}) for $^{6}$Li is shown in arbitrary units as a
function of magnetic field with $\eta=0.001$ and the short range length scale
chosen to be approximately the van der Waals length, $r_{0}=r_{d}\approx30$
a.u. The large y-axis tick marks indicate orders of maginitude. (b) The three
scattering lengths $a^{\left(  1\right)  }$ (solid black curve), $a^{\left(
2\right)  }$ (dashed red curve) and $a^{\left(  3\right)  }$ (dot-dashed blue
curve) are shown in atomic units as a function of magnetic field in Region V.}%
\label{Efiplot}%
\end{figure}

Figure \ref{Efiplot} (a) show the log of the three-body recombination rate in
arbitrary units as a function of the magnetic field with $\eta=0.001$. The
short range length scale here is chosen to be approximately the van der Waals
length of $^{6}$Li, $r_{0}=r_{d}\approx30$ atomic units. Figure \ref{Efiplot}
(b) shows the scattering lengths in the same region of magnetic fields for
reference. An Efimov resonance can clearly be seen at $B=942$ G when
$\phi_{WKB}=3\pi/2$. This is in rough agreement with the predicted position of
$B=1160$ G found in Ref. \cite{braaten2009three}. The exact position of this
resonance is somewhat sensitive to the short-range length scale $r_{0}$ which
should be fit to experimental data. We have chosen $r_{0}$ as the Van der
Waals length here for illustrative purposes. A WKB phase is $3\pi/2$ at the
resonance indicates that this corresponds to the \emph{second} Efimov state
intersecting the continuum. The first Efimov state remains bound throughout
this region. Because $a^{\left(  1\right)  }$ becomes resonantly large as
\thinspace$B\rightarrow834.15$ G, the $\left[  a^{\left(  3\right)  }\right]
^{2}$ scaling from Eq. (\ref{Eq:ScalingR5}) gives the large recombination rate
seen in the lower field region of Fig. \ref{Efiplot} (a).

With three overlapping resonances, $^{6}$Li provides a rich hunting ground for
the study of three-body physics. Further, because it is a Fermionic atom,
three-body interactions involving only two of the three lowest components are
strongly suppressed meaning that the majority of the three-body physics is
controlled by a system of three distinguishable particles. While only the
processes of three-body recombination that lead to trap losses were studied in
this section, there is still a rich and complex array of behaviors not
discussed that can be described using the model presented here.

\section{Summary}

\label{Summary} In this work we have developed a new form for the hyperangular
Green's function in arbitrary dimensions. The derivation of the Green's
function is simple and follows easily from a standard Sturm-Liouville problem.
By dividing a $d$ dimensional space into physically meaningful subspaces, this
new Green's function avoids the slow convergence often seen in spectral
expansions form, while maintaining a physically intuitive set of hyperangular coordinates.

We have also used the hyperangular Green's function to solve the three-body
problem with zero-range s-wave interaction for arbitrary scattering lengths,
particle masses and total angular momentum. With simple root finding, the
adiabatic hyperangular channel functions and adiabatic potentials can be
extracted. The resulting transcendental equation is in exact agreement with
that derived using Fadeev like decompositions. To complete the problem, we
have also derived, for the first time, general expressions for the
non-adiabatic corrections to the potentials that are analytic up to root finding.

The results of the general three-body problem were then applied to the three
lowest hyperfine components of $^{6}$Li near a set of overlapping resonances.
By a simple WKB formalism, the scaling behavior of rate constant for trap loss
three-body recombination events was extracted throughout the overlapping
two-body resonances. Signatures of an Efimov style resonance are also
predicted to appear at high field strengths. Throughout the resonances, all of
the scattering lengths are very large compared to the length scale of the
two-body interaction, indicating that the results presented here are
universal. The simple and intuitive nature of the Lippmann-Schwinger equation
in the three-body problem indicates that this Green's function based method
may be applicable in the context of the four-body problem, but this extension
is the subject of ongoing inquiry.

\section*{Acknowledgements}

The authors would like to thank D. Blume for useful discussions. This research
was supported in part by funding from the National Science foundation. S.T.R.
acknowledges support from a NSF grant to ITAMP at Harvard University and the
Smithsonian Astrophysical Observatory. The authors would like to thank J. P.
D'Incao for many fruitful discussions.


\section*{Appendix A}

\label{appendix} In this appendix we sketch the derivation of the formulas for
the non-adiabatic $\mathbf{P}$ and $\mathbf{Q}$ matrix elements given in Eqs.
~(\ref{Eq:Pmat_elem}) and (\ref{Qmat1}). We begin by considering matrix
elements dealing with the derivative of the Adiabatic Schr\"{o}dinger
equation:%
\begin{align}
\left\langle \Phi_{n}^{\prime}\left\vert \left(  \Lambda^{2}-\varepsilon
_{m}\right)  \right\vert \Phi_{m}\right\rangle  &  =0,\label{deriveq}\\
-\varepsilon_{n}^{\prime}\left\langle \Phi_{m}\left\vert \Phi_{n}\right.
\right\rangle +\left\langle \Phi_{m}\left\vert \left(  \Lambda^{2}%
-\varepsilon_{n}\right)  \right\vert \Phi_{n}^{\prime}\right\rangle  &
=0\nonumber
\end{align}
where $\varepsilon_{n}=\nu_{n}\left(  \nu_{n}+4\right)  $ is the hyperangular
eigenvalue of the $n$th adiabatic eigenfunction, and the prime indicates a
hyperradial derivative has been taken. Taking the difference of these leads to
an equation for the non-adiabatic coupling matrix element $P_{mn}$ for $m\neq
n$:%
\begin{equation}
\left\langle \Phi_{n}^{\prime}\left\vert \Lambda^{2}\right\vert \Phi
_{m}\right\rangle -\left\langle \Phi_{m}\left\vert \Lambda^{2}\right\vert
\Phi_{n}^{\prime}\right\rangle -\left(  \varepsilon_{m}-\varepsilon
_{n}\right)  P_{mn}+\delta_{mn}\varepsilon_{n}^{\prime}=0. \label{Pmateq1}%
\end{equation}
The difference $\left\langle \Phi_{n}^{\prime}\left\vert \Lambda
^{2}\right\vert \Phi_{m}\right\rangle -\left\langle \Phi_{m}\left\vert
\Lambda^{2}\right\vert \Phi_{n}^{\prime}\right\rangle $ is given by the
boundary conditions of the wave functions $\Phi_{m}$ and $\Phi_{n}$ at the
coalescence points:%
\begin{widetext}
\begin{align}
\left\langle \Phi_{n}^{\prime}\left\vert \Lambda^{2}\right\vert \Phi
_{m}\right\rangle -\left\langle \Phi_{m}\left\vert \Lambda^{2}\right\vert
\Phi_{n}^{\prime}\right\rangle  &  =\sum_{k}\left[  \dfrac{a^{\left(
k\right)  }}{d_{k}R}C_{m}\dfrac{\partial}{\partial R}C_{n}^{\left(  k\right)
}-C_{m}^{\left(  k\right)  }\dfrac{\partial}{\partial R}\left(  \dfrac
{a^{\left(  k\right)  }}{d_{k}R}C_{n}^{\left(  k\right)  }\right)  \right]
\nonumber\\
&  =\sum_{k}C_{m}^{\left(  k\right)  }C_{n}^{\left(  k\right)  }%
\dfrac{a^{\left(  k\right)  }}{d_{k}R^{2}}. \label{matdiff}%
\end{align}
\end{widetext}
Here the $LM$ subscript in the boundary values $C_{LM}^{\left(  k\right)  }$
have been suppressed. Inserting Eq.~(\ref{matdiff} into Eq.~(\ref{Pmateq1})
yields Eq.~(\ref{Eq:Pmat_elem}),%
\begin{align}
P_{mn}  &  =\dfrac{\sum_{k}C_{m}^{\left(  k\right)  }C_{n}^{\left(  k\right)
}\dfrac{a^{\left(  k\right)  }}{d_{k}R^{2}}}{\left(  \varepsilon
_{m}-\varepsilon_{n}\right)  }\text{ for }n\neq m\label{Pmat}\\
-\varepsilon_{n}^{\prime}  &  =\sum_{k}\left(  C_{n}^{\left(  k\right)
}\right)  ^{2}\dfrac{a^{\left(  k\right)  }}{d_{k}R^{2}}.\nonumber
\end{align}
A similar derivation provides the matrix elements $Q_{mn}$ given in
Eq.~(\ref{Qmat1}).

\end{document}